\documentclass[fleqn,usenatbib]{mnras}

\usepackage[T1]{fontenc}
\usepackage{ae,aecompl}

\usepackage{graphicx}	
\usepackage{amsmath}	
\usepackage{amssymb}	

\title[The radial SNR distribution in the Galaxy]{The radial supernova remnant distribution in the Galaxy}

\author[Sill Verberne \& Jacco Vink]{
Sill Verberne,$^{1}$
Jacco Vink$^{1,2}$
\\
$^{1}$Anton Pannekoek Institute for Astronomy, University of Amsterdam, 1090 Amsterdam, The Netherlands\\
$^{2}$GRAPPA, University of Amsterdam, 1090 Amsterdam, The Netherlands\\
}

\date{Accepted XXX. Received YYY; in original form ZZZ}

\pubyear{2021}

\begin{document}
\label{firstpage}
\pagerange{\pageref{firstpage}--\pageref{lastpage}}
\maketitle

\begin{abstract}
Supernovae are the dominant source of chemical enrichment of galaxies, and they are an important source of energy to heat the interstellar medium and accelerate cosmic rays. Our knowledge of supernovae in the Milky Way is based mostly on the study of Galactic supernova remnants (SNRs), providing an (incomplete) record to supernova activity over the last $\sim 100,000$~yr. Here we report on an investigation of the spatial distribution of Galactic SNRs.
Given the limited number of SNRs it is common to assume a functional form for the Galactocentric distribution of SNRs. However, several functional forms have been used in the past, without much justification for the radial distribution. For example, one often used functional form implies that no supernova activity is present in the Galactic Centre region. However, the presence of a magnetar and a SNR near the Galactic Centre suggest that a spatial distribution with zero SNRs at the Galactic Centre is not realistic.
In light of these concerns we reevaluate the Galactic SNR distribution. We provide a brief outline of the main detection biases in finding SNRs and we investigate whether or not the use of the most common functional form is justified and how it compares to other models for the SNR distribution. We do this by analysing the longitudinal distribution of SNRs. 
We find that a simple exponential distribution is the most consistent and simplest model for describing the radial SNR distribution in the Galaxy and draw comparisons with the massive star formation and metallicity distributions.

\end{abstract}

\begin{keywords}
ISM: supernova remnants -- supernovae: general -- Galaxy: structure
\end{keywords}



\section{Introduction}
Supernova remnants (SNRs) are formed by both core-collapse and thermonuclear supernovae (SNe). Core-collapse SNe are associated with massive stars at the end of their life \citep[e.g.][]{Woosley_1995, Bethe_1990} and thermonuclear SNe are associated with the explosions of CO white dwarfs \citep[e.g.][]{Hillebrandt_2000, Moaz_2014}. SNe occur at a rate of about two to three per century in the Galaxy \citep[e.g.][]{Li_2011, Tammann_1994}. The SNRs formed by these SNe are believed to remain visible for 20 to 80 kyr \citep{Sarbadhicary_2017}. Given the rate at which they occur and their expected lifetime, only a fraction of the SNRs in the Galaxy have actually been observed. Furthermore, the sample of SNRs that have been discovered is heavily biased by selection effects such as distance and surface brightness. 

SNRs are of interest to the study of cosmic rays since they are believed to be sites of cosmic-ray acceleration up to PeV energies \citep[e.g.][]{Blasi_2011, Berezhko_2003, Bell_2013}. Furthermore, they can provide physical insight into the physics of SNe themselves. SNe are the most important source of alpha-elements and Fe-group elements for enriching the interstellar medium, which can be studied in SNRs. Since $\sim80\%$ of SNe are of the core collapse type \citep[e.g.][]{Li_2011, Graur_2017} which only occur for massive stars with relatively short lifetimes, the spatial SNR distribution is expected to follow massive-star formation. A good model for the spatial distribution would allow more focused searching for new remnants by comparing the model with the observed SNRs and predicting the location of the highest number of undiscovered remnants. This could in turn result in a better understanding of SNe themselves, since SNRs contain information about the SNe that formed them.

\citet{Green_2015} relied on the 1D projection, Galactic longitude ($l$), of the spatial distribution, whereas \citet{Case_1998} attempted to reconstruct the full spatial distribution by using the $\Sigma-D$ relation to infer distances to all SNRs.

The aim of this work is to expand upon the study of \citet{Green_2015} by fitting and comparing multiple functional forms. In earlier work, the use of one of many possible functional forms seems to suggest there is only one possible description of the radial SNR distribution. The problem with the model usually adopted is that it approaches $0$ at the Galactic Centre, while SNRs are known to reside in this region as described by \citet{Maeda_2002} and \citet{Kennea_2013}. Whereas \citet{Green_2015} fitted to minimize the residuals of the cumulative distribution function, we will fit directly to the observed number of SNRs per bin in $l$.

In section \ref{sec:SNR detection biases} the most important biases in the SNR catalogue will be outlined. Section \ref{sec:methods} discusses the methods involved in the investigation. In section \ref{sec:fits} the results of the performed model fits will be given and sections \ref{sec:discussion} and \ref{sec:conclusion} will give the discussion and conclusions respectively. 


\section{Supernova remnant detection biases}
\label{sec:SNR detection biases}
This work is based upon Green's Catalogue of Galactic SNRs \citep{Green_2019} which contains 294 SNRs. In this section an outline will be given of the most important selection effects in the discovery of these SNRs.

\citet{Green_2004, Green_2015} advocates for the use of a nominal surface-brightness ($\Sigma$) completeness limit of $10^{-20}$ W m$^{-2}$ Hz$^{-1}$ sr$^{-1}$, which is a conserved quantity since
\begin{equation}
    \Sigma = \frac{F}{\Omega}; \hspace{1cm} F=\frac{L}{4\pi R^2}; \hspace{1cm} \Omega = \frac{A}{R^2},
\end{equation}
with $F$ the flux density, $L$ the luminosity, $\Omega$ the solid angle, $A$ the projected surface area, and $R$ the distance to the source. The reason for imposing a surface-brightness threshold in the radio is that most remnants have been discovered in the radio and surface brightness is an important selection effect in the discovery of new remnants. Imposing the threshold thus decreases the bias of the sample, although, as mentioned by \citet{Green_2015}, a higher surface-brightness threshold might be required around the Galactic Centre due to higher Galactic background radio emission. Furthermore, the sample also suffers from a bias disfavouring small angular-sized remnants, since they might be too small to be properly identified. This problem with small angular-sized remnants is not resolved by imposing a surface-brightness threshold. What we do know is that these small angular size remnants will be concentrated towards small $l$ (or large physical distances). Moreover, due to source confusion along the line of sight, even SNRs of sufficient size are more difficult to detect near the Galactic Centre. This means that both the surface brightness, small angular size, and confusion biases are most important around the Galactic Centre. 

Bias in the sample selection could also result from the correlation between the surface brightness of SNRs and the density of the medium in which they evolve. It is believed that for a remnant evolving in a relatively high density medium its surface brightness will be high and its lifetime low compared to one evolving in a relatively low density environment \citep[e.g.][]{Blondin_1998, Sarbadhicary_2017}. Fortunately for this study, these effects work to counteract each other; although a SNR evolving in a high density environment is more likely to exceed a surface-brightness-completeness limit it will be relatively short lived. We thus expect the net contribution of this bias to be relatively minor.

\subsection{Unbiased sample selection}
In order to investigate the bias introduced by the higher background around the Galactic Centre we use the 408 MHz all sky radio map from \citet{Haslam_1982}. In order for a SNR to be recognized as such, the surface brightness needs to exceed the noise level of the background sufficiently. Using only the background level for this purpose does not capture the full complexity by which SNRs are discovered (e.g.~remnants are easier to discover from multi-frequency observations by decomposing the thermal and non-thermal components). Also, it does not alleviate the bias induced by small angular size remnants or source confusion along the line of sight. However, it does give us the order of magnitude by which the SNR completeness limit of $10^{-20}$ W m$^{-2}$ Hz$^{-1}$ sr$^{-1}$ is off in the region around the Galactic Centre. By adjusting the surface-brightness cutoff we can thus limit the dominant sources of bias to small angular sizes and source confusion along the line of sight.\

\begin{figure}
    \centering
    \includegraphics[width=0.47\textwidth]{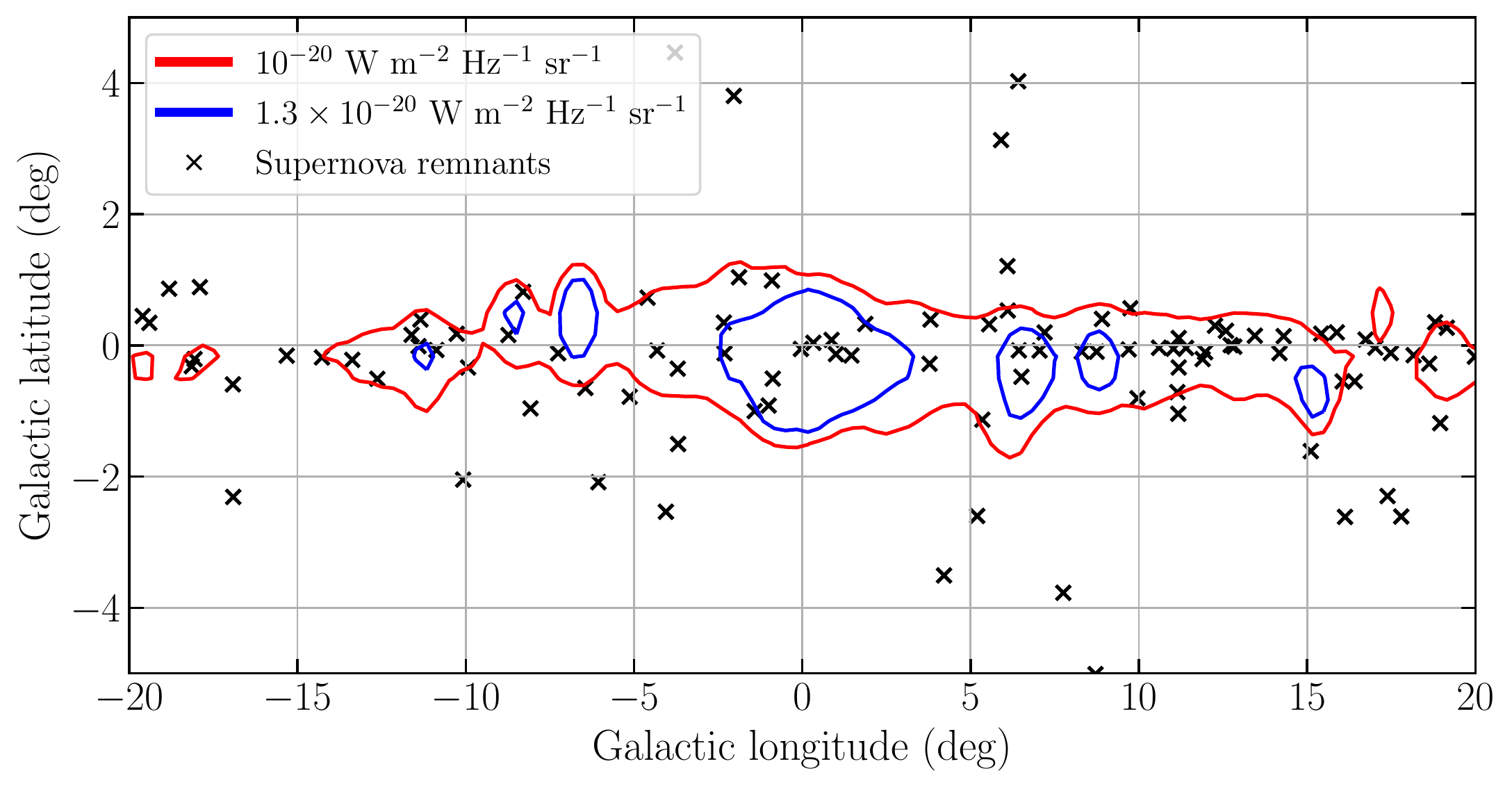}
    \caption{Contours of the surface-brightness threshold values of $10^{-20}$ W m$^{-2}$ Hz$^{-1}$ sr$^{-1}$ and $1.3 \times 10^{-20}$ W m$^{-2}$ Hz$^{-1}$ sr$^{-1}$ at 1 GHz are shown near the Galactic Centre along with all SNRs discovered in the region.}
    \label{fig:contours}
\end{figure}

\begin{figure}
    \centering
    \includegraphics[width = 0.47\textwidth]{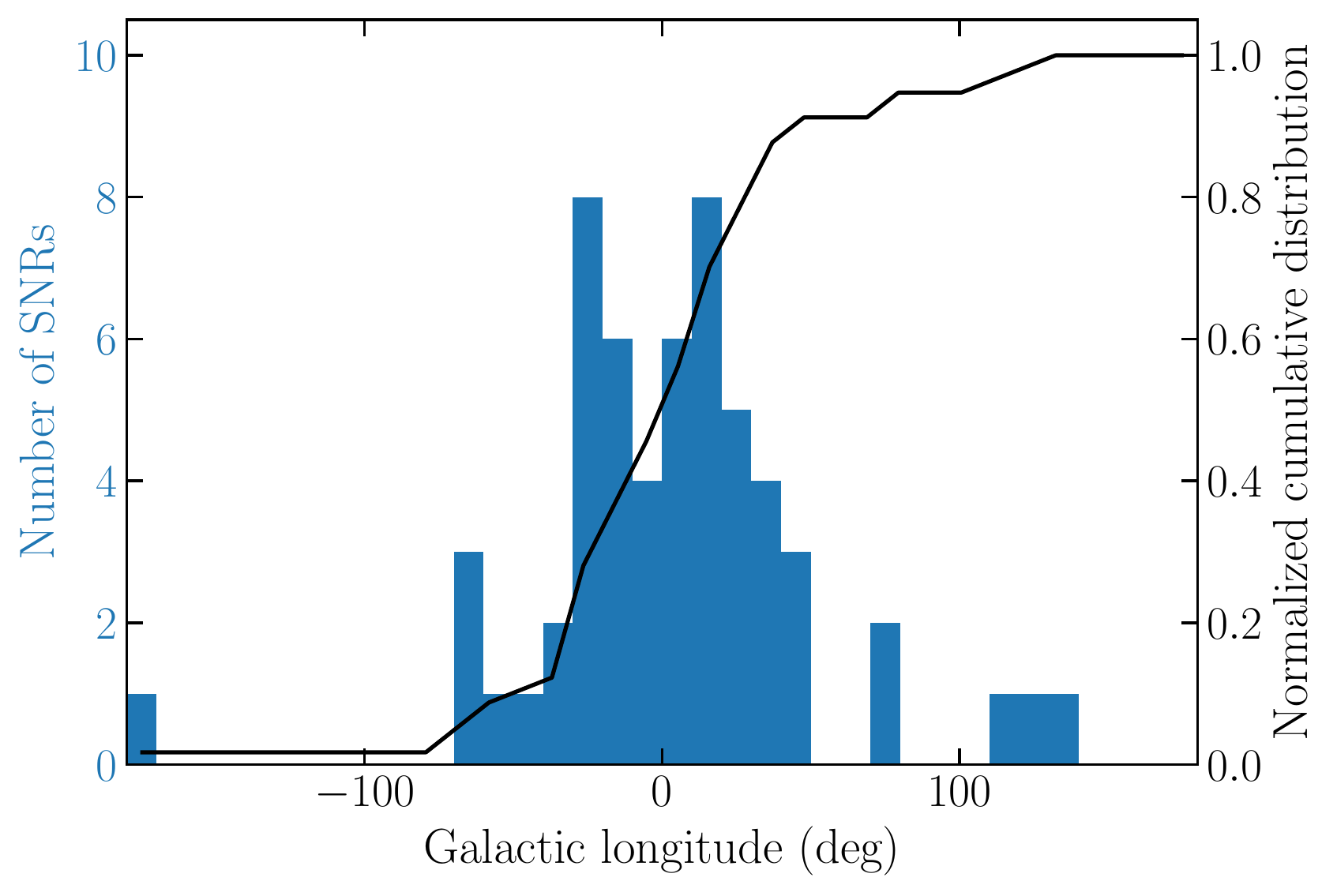}
    \caption{Histogram showing the distribution of SNRs in longitude along with the normalized cumulative distribution. Only the SNRs that pass the threshold at 1 GHz of $1.3 \times 10^{-20}$ W m$^{-2}$ Hz$^{-1}$ sr$^{-1}$ have been included here.}
    \label{fig:hist_l}
\end{figure}

Since the radio sky is dominated by synchrotron emission at 408 MHZ, we can use its spectral index to extrapolate the map to 1 GHz using
\begin{equation}
    \textrm{T}(\nu) \propto \nu^{-\beta},
\end{equation}
with $T(\nu)$ the brightness temperature, $\nu$ the frequency and $\beta$ the spectral index. \citet{Platania_1998} determined that the spectral index for synchrotron radiation is $2.76\pm0.11$ in the range $0.4-7.5$ GHz. Using this spectral index, while adjusting for the cosmic microwave background, we created a radio map at 1 GHz. We then investigated the remaining bias around the Galactic Centre caused by using a surface-brightness threshold of $10^{-20}$ W m$^{-2}$ Hz$^{-1}$ sr$^{-1}$ by making a contour plot. In Fig.~\ref{fig:contours} this plot is shown. It can be seen that the $10^{-20}$ W m$^{-2}$ Hz$^{-1}$ sr$^{-1}$ surface-brightness threshold is not high enough to prevent bias for $|l|\lesssim 20^\circ$ due to the high background. 
By increasing the surface-brightness threshold to $1.3 \times 10^{-20}$ W m$^{-2}$ Hz$^{-1}$ sr$^{-1}$ we can reduce this biased region to $|l|\lesssim 10^\circ$, as shown by the blue contours in Fig.~\ref{fig:contours}. In this region we know that there is bias anyway due to the small angular sizes and source confusion. We have thus effectively limited the biased region to this inner $\pm10^\circ$ by increasing the surface-brightness threshold. In section~\ref{sec:methods} we will discuss how this remaining bias is addressed.
Of the 294 SNRs in the Catalogue, 67 pass the threshold of $10^{-20}$ W m$^{-2}$ Hz$^{-1}$ sr$^{-1}$ at 1 GHz and 57 that of $1.3 \times 10^{-20}$ W m$^{-2}$ Hz$^{-1}$ sr$^{-1}$ at 1 GHz. In Fig.~\ref{fig:hist_l} a histogram of the distribution in $l$ is shown of the remnants that pass the $1.3 \times 10^{-20}$ W m$^{-2}$ Hz$^{-1}$ sr$^{-1}$ surface-brightness threshold at 1 GHz along with their cumulative distribution. The $1.3 \times 10^{-20}$ W m$^{-2}$ Hz$^{-1}$ sr$^{-1}$ surface-brightness threshold at 1 GHz is what will be used for the remainder of this paper.

\section{Methods}\label{sec:methods}
\subsection{Analysis}\label{sec:analysis}
\begin{figure}
    \centering
    \includegraphics[width=0.47\textwidth]{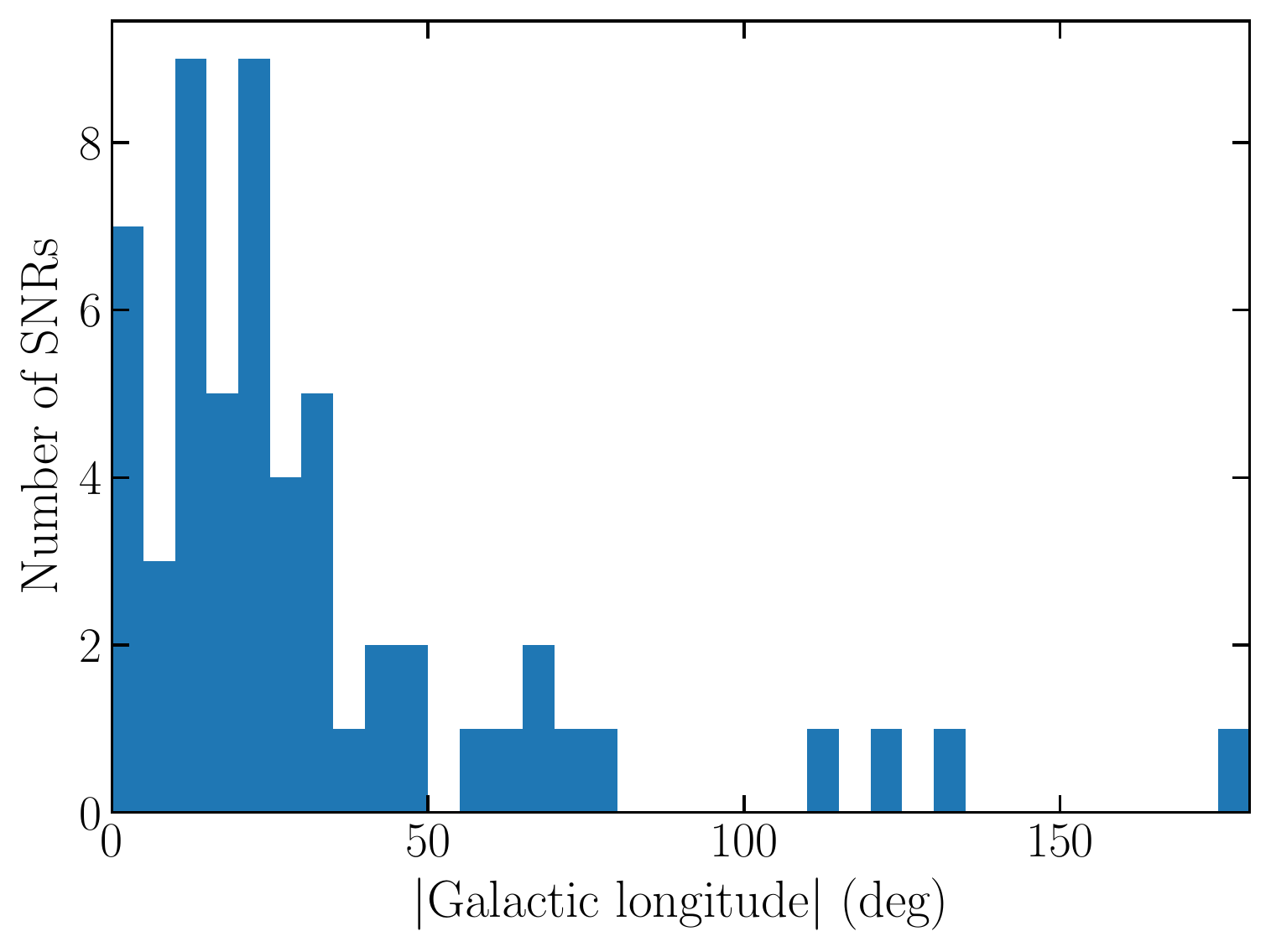}
    \caption{The SNR distribution as a function of the absolute value of $l$. Only SNRs satisfying the surface-brightness threshold of $1.3 \times 10^{-20}$ W m$^{-2}$ Hz$^{-1}$ sr$^{-1}$ at 1 GHz have been included.}
    \label{fig:fold}
\end{figure}

We follow \citet{Green_2015} in using the 1D projection of the SNR distribution in $l$ instead of the 3D $\Sigma - D$ relation \citep[e.g.][]{Case_1998}. The reason for this is that the $\Sigma - D$ relation is an empirical relation between the surface brightness ($\Sigma$) and the diameter of the source ($D$), that suffers from scatter of about an order of magnitude in the derived distance to said source. The $l$ of SNRs is not subject to this problem and can still be compared with model distributions. Since we only consider cylindrically symmetric distributions, the absolute value of $l$ has been taken in order to improve the statistics per bin; 36 bins of $5^{\circ}$ each between $0$ and $180^\circ$ were used. In Fig.~\ref{fig:fold} the resulting histogram of the surface-brightness-limited sample is shown.

Because the fitted distributions are functions of the Galactocentric radius ($R_{\rm{gal}}$), which we do not directly observe, a conversion to a heliocentric radius ($R_{\rm{helio}}$) needs to be made. This conversion can be made using the law of cosines as
\begin{equation}
    R_{\rm{gal}} = \Big(R_{\rm{0}}^{2} + R_{\rm{helio}}^{2} - 2R_{\rm{0}}R_{\rm{helio}}\cos{l}\Big)^{1/2},
\end{equation}
where $R_0$ is the distance to the centre of the Galaxy of about $8$ kpc \citep{Genzel_2010} and $l$ the Galactic longitude. After this conversion, Simpson's rule of numerical integration has been used in the range $0 \leq R_{gal} < 50$ kpc in order to obtain the number of SNRs along the line of sight in a given direction according to the model distribution. This model distribution is then fit to the observed distribution using the $\chi^2$ statistic 
\begin{equation}
\label{eq:chi2}
    \chi^2 = \sum_{\rm{i}=1}^{N}\frac{(n_{\rm{i}} - e_{\rm{i}})^2}{\sigma_{\rm{i}}^2},
\end{equation}
with $n_{\rm{i}}$ the observed counts per bin, $e_{\rm{i}}$ the model counts per bin and $\sigma_{\rm{i}}$ the error on the observed counts per bin.

Since the SNR Catalogue with the imposed surface-brightness-completeness limit only has 57 remnants, the Catalogue is a sparse data set, which means that $\sigma_{\rm{i}}$ cannot be approximated from the observed counts per bin. Instead we need to use Poisson statistics to estimate $\sigma_{\rm{i}} = \sqrt{e_{\rm{i}}}$. \citet{Wheaton_1995} note that in order to avoid biases, the weighting factors (i.e. $\sigma_{\rm{i}}$'s) need to be taken as constants before the fitting process. This fitting process can then be iterated over while adjusting $\sigma_{\rm{i}}$ from run to run. This results in a model fit with unbiased parameters even when $n_{\rm{i}}\ll1$. In order to convince the critical reader of its effectiveness, fitting has been repeated directly using the Poisson distribution (maximum log-likelihood). Parameter errors have been determined by grid searches where (using $\chi^2$ minimization) the $1\sigma$ confidence limit coincides with where $\chi^2 = \chi^2_{\rm{min}}+1$. In these grid searches, the errors (i.e. $\sigma_{\rm{i}}$) where based on the best fitting model.

As a goodness of fit test, the $Y^2$ statistic \citep{Lucy_2000} has been used. This is, like Pearson's statistic, a $\chi^2$ statistic with the added benefit of being more accurate when the amount of counts per bin is low. In fact, the statistic remains reasonably accurate at the $2\sigma$ level when the mean counts per bin are $\ll1$ as long as the total number of events is $\gtrsim 30$ \citep{Lucy_2000}. The $Y^2$ statistic, in the multinomial case, is given by
\begin{equation}
\label{eq:Y2}
    Y^2 = \nu + \xi^{-1}(X^2-\nu); \hspace{1cm} \xi^{-1} = \sqrt{\frac{2\nu}{V(X^2)}},
\end{equation}
where $\nu$ is the number of degrees of freedom and 
\begin{equation}
    V(X^2) = 2\nu + \sum_i\frac{1}{e_{\rm{i}}} - \frac{1}{N}(I^2+2I-2); \hspace{0.5cm} X^2=\sum_{i=1}^{N}\frac{(n_{\rm{i}} - e_{\rm{i}})^2}{e_{\rm{i}}},
\end{equation}
with $I$ being the number of bins used and $N$ the total number of SNRs in the sample. The multinomial case is used since the model only describes the distribution of SNRs and not the number of them, meaning that the number of predicted SNRs is simply the number of observed ones ($\Bar{N}=N$).

Given the relatively large bias in the data set around the Galactic Centre, the model fits have also been performed when omitting the inner $\pm10^\circ$ (2 bins) of SNRs. What this allows us to do is fit the distributions to a data set with less observational bias. However, note that the fits performed on this modified data set should not be used to infer properties of the inner $\pm10^\circ$ of the Galactic SNR distribution.

\subsection{Functional forms}
\label{sec:funcforms}
A number of distributions have been fitted using equation~(\ref{eq:chi2}) to see which one best describes the data. Since the distributions are normalized to the data, no normalization parameters have been included.

In earlier work \citep[see][]{Stecker_1977, Case_1998, Green_2015} there has been an insistence on the use of a distribution of the form
\begin{equation}
\label{eq:gamma_distribution}
    f(r) = \Bigg(\frac{r}{r_\odot}\Bigg)^\alpha \textrm{exp}\Bigg(-\beta\frac{r-r_\odot}{r_\odot}\Bigg),
\end{equation}
which is a modified version of a gamma distribution (henceforth referred to as MGD) with $\alpha$ and $\beta$ being free parameters and $r_\odot$ the distance to the centre of the Galaxy. No attempts have so far been made to see if this distribution is appropriate in this case and no real justification has been given for its use. Therefore, in order to see if this distribution is truly fit for describing the SNR distribution, model fits have also been performed with simple power-law (PL) and exponential distributions of the forms:
\begin{equation}
\label{eq:pl}
    f(r) = \Bigg(\frac{r}{r_\odot}\Bigg)^{\alpha}
\end{equation}
and
\begin{equation}
\label{eq:exp}
    f(r) = \textrm{exp}\Bigg(-\beta\frac{r-r_\odot}{r_\odot}\Bigg),
\end{equation}
where both functions have a single free parameter. The benefit of looking at these two distributions is that we can easily compare them to equation~(\ref{eq:gamma_distribution}) since they are nested models.

A S\'ersic profile \citep{Sersic_1963}, which is a generalization of a de Vaucouleurs profile \citep{deVaucouleurs_1948}, has also been included in the analysis. The S\'ersic profile can be used to describe the stellar or intensity distribution in a galaxy and is given by
\begin{equation}\label{eq:sersic}
    f(r) = \rm{exp}\Bigg\{-b_{\rm{n}}\Bigg[\Bigg(\frac{r}{r_{\rm{e}}}\Bigg)^{1/n}-1\Bigg]\Bigg\},
\end{equation}
where $n$ is a free parameter, $r_{\rm{e}}$ the half-light radius of the galaxy and $b_{\rm{n}}$ a constant that can be approximated by
\begin{equation}
    b_{\rm{n}} \simeq 1.9992n - 0.3271
\end{equation}
for $0.5 < n < 10$. An extensive discussion on this profile can be found in \citet{Graham_2005}. This profile has an additional constraint in comparison to the exponential, since we need the half-light radius which is not treated as a free parameter and assumed to be 3 kpc.

Finally, a sum between an exponential and PL has been included in the form of 
\begin{equation}\label{eq:EXP+PL}
    f(r) = \Bigg(\frac{r}{r_\odot}\Bigg)^{\alpha} + \textrm{exp}\Bigg(-\beta \frac{r-r_\odot}{r_\odot}\Bigg).
\end{equation}
This gives us a set of five models that are combinations between PLs and exponentials that will be used to determine which one best describes the radial SNR distribution in the Galaxy.

Of course one could imagine many more possible models to fit to the data, but in doing so the chances of finding a model that purely by coincidence fits very well would also increase.


\section{Results}\label{sec:fits}
The fitting results of the models to the surface-brightness-limited sample are shown in Table~\ref{tab:params}. We list here a complete sample in the range $0^\circ\leq |l| \leq 180^\circ$ and a centre-excluded sample in the range $10^\circ \leq |l| \leq 180^\circ$. The separate model fits are all shown in Appendix \ref{sec:Plots}. Of the 57 remnants with a surface brightness at 1 GHz of at least $1.3 \times 10^{-20}$ W m$^{-2}$ Hz$^{-1}$ sr$^{-1}$, 47 are outside of the inner $\pm10^\circ$ of the Galaxy. As mentioned in section \ref{sec:analysis}, the fits have also been performed directly using the Poisson distribution. The results (i.e. the fitted parameters) from using the Poisson distribution were exactly equal in the reported significance to those obtained using the $\chi^2$ statistic with iterations over the errors. Note that for models that include a power-law component, this component is not allowed to become smaller than $-1$, since this would result in an infinite number of predicted SNRs at the Galactic Centre. Table~\ref{tab:nested} gives the nested model significances between the MGD and PL, and exponential models. A small significance ($<0.05$) indicates that the MGD fits significantly better, while a large significance ($>0.05$) indicates that it does not.

\begin{table*}
	\centering
	\def\arraystretch{2.0}
	\caption{Parameters and goodness of fit values for the fits performed with the models described in section \ref{sec:funcforms}. All SNRs with a surface brightness above $1.3 \times 10^{-20}$ W m$^{-2}$ Hz$^{-1}$ sr$^{-1}$ at 1 GHz have been used in this analysis. The peak refers to the radius at which the highest number of SNRs is predicted.}
	\label{tab:params}
	\begin{tabular*}{0.9\textwidth}{llccclcc}
	    \hline
	    &\multicolumn{4}{  c  }{Complete sample}& \multicolumn{3}{  c  }{Sample excluding inner $\pm10^\circ$}\\
		\hline
		\textbf{Model} & \textbf{Parameters} & \textbf{Peak (kpc)}& \textbf{$Y^2$} & \textbf{P-value} &\textbf{Parameters} & \textbf{$Y^2$} & \textbf{P-value}\\
		\hline
		MGD (Eq.\ref{eq:gamma_distribution})& $\alpha = 0.74_{-0.91}^{+1.38}$ & 3.9& 33.65 & 0.44 & $\alpha = -0.96_{-0.04}^{+0.38}$& 28.03& 0.62\\ 
		                   & $\beta = 3.52_{-1.51}^{+2.43}$&        & &      & $\beta = 1.71_{-1.58}^{+2.17}$&&\\
		\hline
		Exponential + PL (Eq.\ref{eq:EXP+PL})& $\alpha = -1.00_{-0.00}^{+0.41}$ &0 &28.89 & 0.67 & $\alpha = -0.91_{-0.09}^{+0.61}$& 25.80& 0.73\\ 
		                        & $\beta = 2.90_{-0.67}^{+0.54}$&      &    &      & $\beta = 3.70_{-0.91}^{+0.95}$&&\\
		\hline
		Exponential (Eq.\ref{eq:exp})& $\beta = 2.46_{-0.33}^{+0.39}$&3.3 & 31.82& 0.57 & $\beta = 2.88_{-0.46}^{+0.56}$& 31.13& 0.51\\
		\hline
		PL (Eq.\ref{eq:pl})& $\beta = -1.00_{-0.00}^{+0.03}$& - & 58.02& 0.006 & $\beta = -1.00_{-0.00}^{+0.04}$& 59.27& 0.002\\
		\hline
		S\'ersic profile (Eq.\ref{eq:sersic})& $n = 1.95_{-0.67}^{+0.84}$&0.9 &51.25& 0.03 & $n= 2.56_{-0.75}^{+1.40}$& 27.39& 0.70\\
		\hline
	\end{tabular*}
\end{table*}

\subsection{Modified gamma distribution}
The model fits of the MGD from equation~(\ref{eq:gamma_distribution}) show good fits for both the SNR samples with p-values of 0.44 and 0.62 respectively. For the complete sample fit, the distribution goes to 0 at the Galactic Centre and has its surface density peak at about 1.7 kpc from the Galactic Centre. A more meaningful number, however, is the peak of the number of SNRs at a given radius. In this way we need to also consider the area $2\pi r\Delta r$ of a ring at radius $r$. The highest number of SNRs are expected at a Galactocentric radius of 3.9 kpc.

\subsection{Exponential + power-law distribution}
The distribution made up of the sum of an exponential and a PL gives similar results between the two samples with p-values of 0.67 and 0.73, indicating that this model again fits the data well. Contrary to the MGD, this distribution does not go to 0 near the Galactic Centre. Instead it goes to $\infty$ since $f(r)\propto r^{-1.00}$ in the PL component. As mentioned before, for a power-law component $<-1$ the number of expected SNRs would go to infinity at the Galactic Centre. To avoid this, we forced $\alpha>-1$. This is also the reason that there is no negative error on this value. For the complete surface-brightness-limited sample the best fitting power-law component is $-1$. This means that when integrating over the surface the power law falls out and, since we are then left with an exponential, the maximum number of SNRs is expected at $R_{gal}=0$. 

\subsection{Exponential distribution}
The exponential distribution again fits both samples well with p-values of 0.57 and 0.51 for the complete and centre-excluded samples respectively. The peak of the distribution is at the Galactic Centre but one again needs to realize that the surface area at a given radius plays a role as well. This means that the maximum number of SNRs at a given radius is at about 3.3 kpc.

\subsection{Power-law distribution}
The PL model fits both samples poorly (p-values $\ll0.05$). The problem with this model is that, just like with the sum of an exponential and PL model, the distribution goes to infinity at the Galactic Centre. In this case however, this constraint on the power law is more limiting to the model due to it being the only free parameter.

\subsection{S\'ersic profile}
In the complete SNR sample set, the S\'ersic profile provides a poor fit with a p-value of 0.03 and a peak in the expected number of SNRs at 0.9 kpc. However, a large improvement can be seen when moving to the centre-excluded model where we find a p-value of 0.70. Looking at Appendix~\ref{fig:compl}, we see that the poor fit for the complete sample is likely due to the model distribution becoming very small at large angles. 

\subsection{Nested models}
The values reported in Table~\ref{tab:nested} show what the significances are when comparing how well the shown distributions fit. The difference in their $Y^2$ values can be compared using that it is distributed as a $\chi^2_\nu$ distribution with $\nu$ the difference in the number of degrees of freedom between the models, since by going from the MGD to either the PL or exponential model a degree of freedom effectively gets added. The value of 1 implies that the model with the larger amount of degrees of freedom (i.e. the exponential) fits better than the one with fewer (i.e. the MGD). In section~\ref{sec:discussion} this will be discussed.

\begin{table}
    \centering
    \caption{The nested-model significances of the MGD and exponential, and power-law distributions for both the complete and centre-excluded samples.}
    \label{tab:nested}
    \begin{tabular*}{0.49\textwidth}{lcc}
    \hline
    & Complete sample & Centre-excluded\\
    & significance & sample significance\\
    \hline
    MGD \& Exponential& 1.0 & 0.078\\
    MGD \& PL& $8\times10^{-7}$ & $2\times10^{-8}$\\
    \hline
    \end{tabular*}
\end{table}

\section{Discussion}\label{sec:discussion}
One of the caveats of this investigation is that it is not clear from a theoretical point of view what the SNR distribution should look like. As discussed in the introduction, we in general expect it to follow the massive star formation. But a complicating factor are type Ia SNe which cannot be expected to follow massive star formation in any trivial sense. For this reason, the models investigated here were chosen as to have a mix of distributions and investigate if the use of the MGD from equation~(\ref{eq:gamma_distribution}) is really appropriate. 

Instead of the $10^{-20}$ W m$^{-2}$ Hz$^{-1}$ sr$^{-1}$ surface-brightness-completeness limit that \citet{Green_2004, Green_2015} use, we opted for a slightly higher limit of $1.3 \times 10^{-20}$ W m$^{-2}$ Hz$^{-1}$ sr$^{-1}$ based upon the Galactic background near the Galactic Centre. Using this surface-brightness-limited sample we effectively limited the bias in the SNR sample to the inner $\pm10^{\circ}$ of the Galaxy.

\citet{Green_2015} used the least sum of squares of the differences between the observed and model cumulative distributions. Since in the cumulative distribution the bins are not independent, we instead use the $\chi^2$ statistic from equation~(\ref{eq:chi2}) to compare the observed and model number of SNRs per bin in $l$. This method was chosen in order to avoid biases from using unweighted fitting to data bins that are not independent. The fact that the minimization of the $\chi^2$ statistic with iterations over $\sigma_{\rm{i}}$ provided equal results to fits performed with the use of the Poisson distribution shows that the parameter estimates are reliable.

The SNR catalogue has been updated since \citet{Green_2015}'s investigation and along with the increased surface-brightness threshold this has resulted in twelve remnants fewer in the sample used here compared to that used in \citet{Green_2015}. In combination with the improved fitting presented here this causes the distribution to be more centrally concentrated than the one found by \citet{Green_2015}. We find that the MGD fit of the surface density peaks at a radius of 1.68 kpc while \citet{Green_2015} and \citet{Case_1998} find values of 2.25 and 4.53 kpc respectively. Like \citet{Green_2015} noted, the relatively large difference with the distribution found by \citet{Case_1998} can be explained by their use of the $\Sigma - D$ relation which causes bias in their sample.

\subsection{Model discussion}
The results shown in Table~\ref{tab:params} indicate that neither in the complete nor centre-excluded samples, fitting a MGD provides significantly better fits than a simple exponential distribution. Although the MGD does come close to being significantly better in the centre-excluded one, it is in the complete sample where the exponential actually performed better while having less free parameters. The reason this can happen is that the iterations over the errors that were fitted make the parameters deviate from the global minimum of the $\chi^2$ statistic. What we also see in the MGD fit for the complete sample is that the distribution in $l$ does not have its peak at $0^\circ$. In section \ref{sec:SNR detection biases} it was explained that it is likely that not all SNRs with a surface brightness above our threshold of $1.3 \times 10^{-20}$ W m$^{-2}$ Hz$^{-1}$ sr$^{-1}$ at 1 GHz have been found near the Galactic Centre. The drop in the number of SNRs predicted by the MGD below about $\pm10^{\circ}$ could thus be due to bias in the data. However, this bias is also present for the other fitted models which do not show this feature. This makes it interesting to investigate if this feature remains present when a more complete SNR sample is analysed.

Apart from the statistical significance argument, there is another important observational argument to be made against the use of the MGD. While the MGD does not predict SNRs near the Galactic Centre, observations \citep[e.g.][]{Maeda_2002, Kennea_2013} have shown SNe to take place near the centre of the Galaxy thus causing a mismatch between the model and observations. 

The reason that we have focused mainly on the MGD and exponential profiles is that the PL provides poor fits to both samples. Furthermore, the exponential + PL model cannot be compared in the same way since the PL and exponential distributions are not nested models of it. Moreover, the predicted maximum in the number of SNRs from this model is at the Galactic Centre itself, which we know to be untrue. The S\'ersic profile also does not fit the complete sample well and we find that the S\'ersic index $n$ for the complete sample is higher than the $1.3\pm0.3$ found by \citet{Widrow_2008}. When going to the sample with the inner $\pm10^\circ$ excluded we find an even higher number for $n$ that would correspond to a bulge dominated system \citep{Driver_2006}. In combination with the poor fit to the complete sample we can quite safely say that the Galactic SNR distribution does not follow a S\'ersic profile.

\begin{figure}
    \centering
    \includegraphics[width=0.47\textwidth]{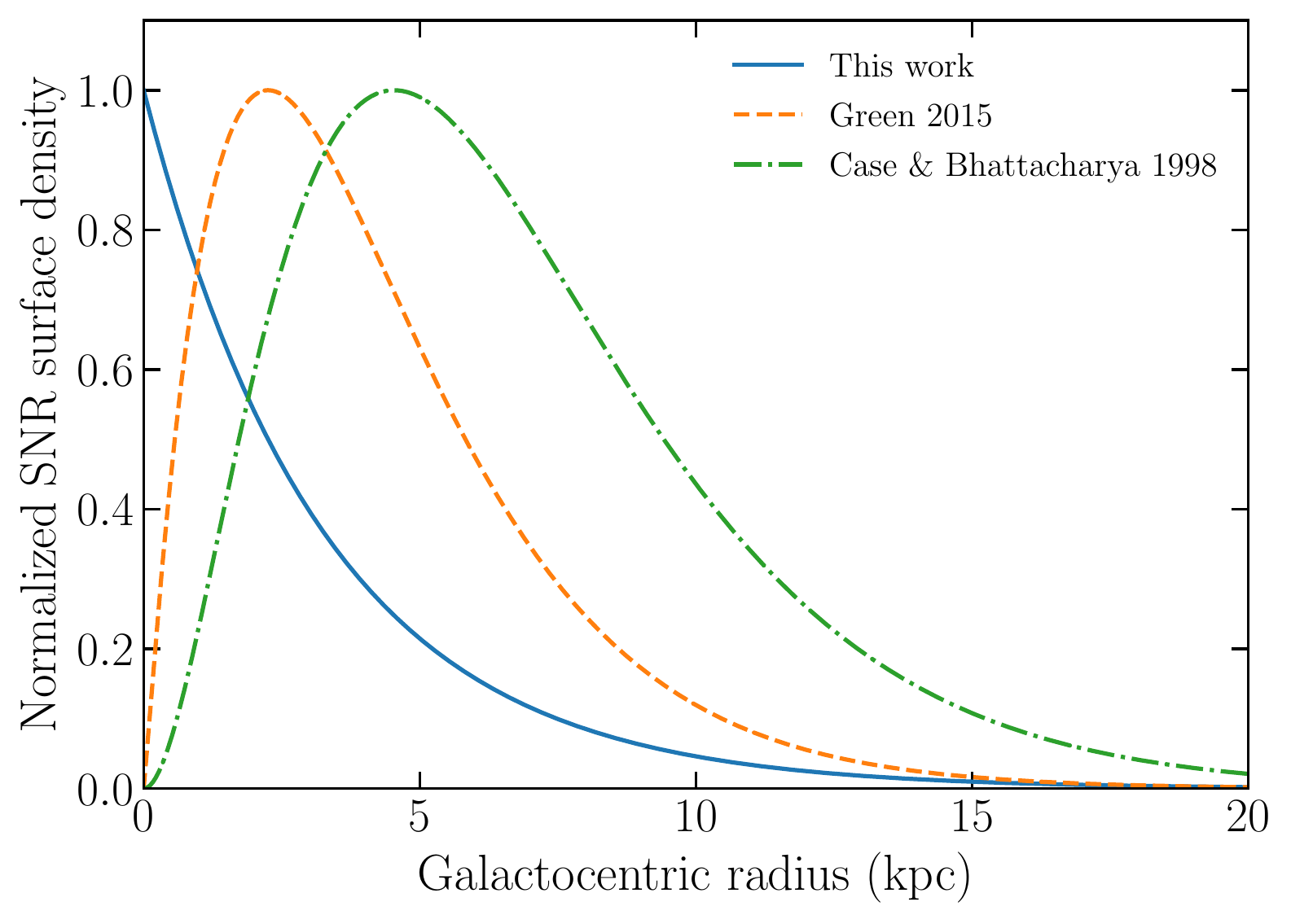}
    \caption{The best fit normalized surface density of the radial SNR distribution in the Galaxy as determined by this work in comparison with those found by \citet{Case_1998} and \citet{Green_2015}. Here we use the parameters obtained from the complete surface-brightness-limited sample.}
    \label{fig:comparison_dens}
\end{figure}

\begin{figure}
    \centering
    \includegraphics[width=0.47\textwidth]{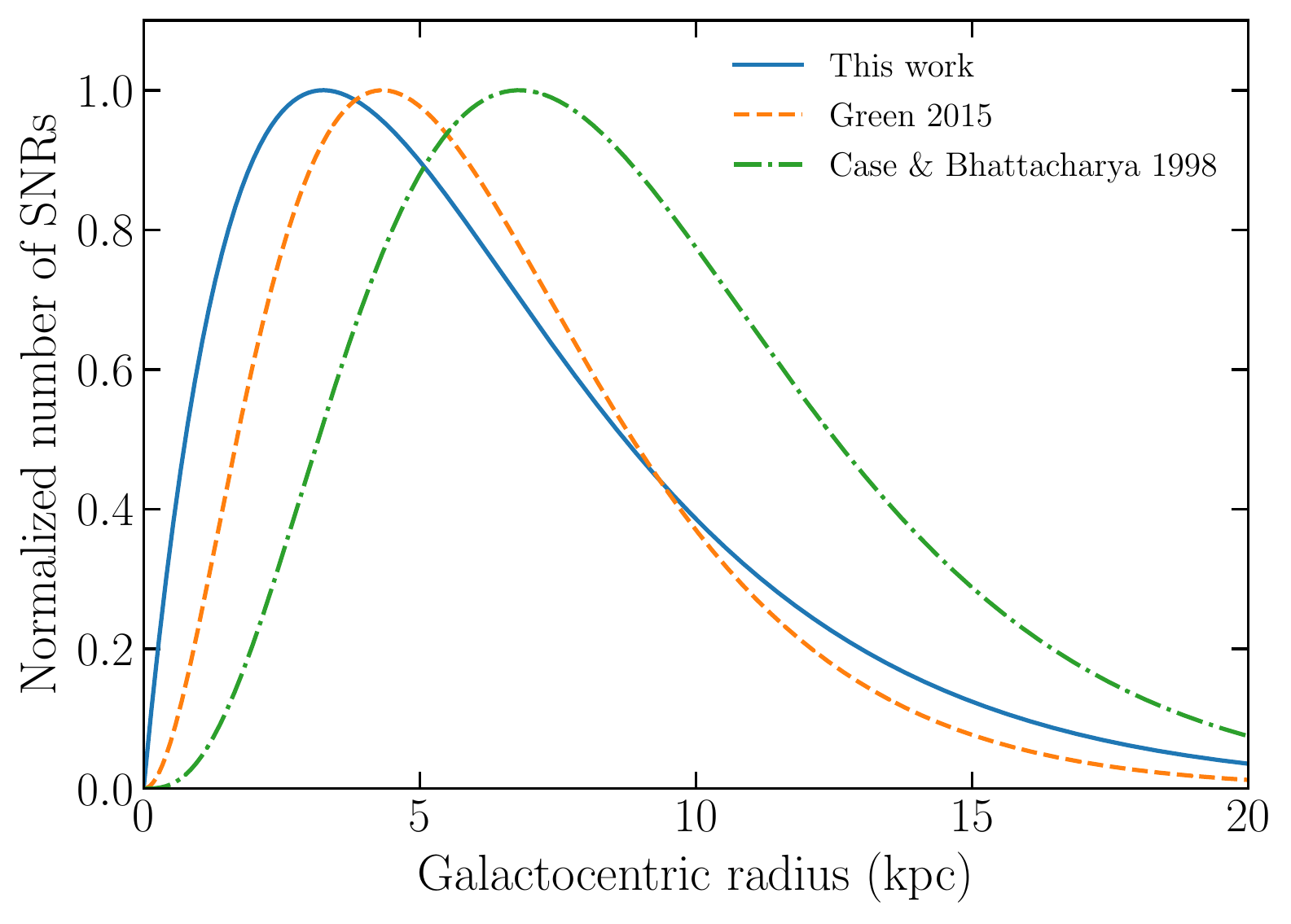}
    \caption{Same as Fig.~\ref{fig:comparison_dens}, but here we show the normalized number density (integrated surface density) instead.}
    \label{fig:comparison_num}
\end{figure}

From the above discussion we conclude that the simple exponential distribution provides the best fit to the data. In Fig.~\ref{fig:comparison_dens} and Fig.~\ref{fig:comparison_num} we, therefore, show a normalized surface density and number density comparison respectively between this distribution and the best fit MGDs found by \citet{Case_1998} and \citet{Green_2015}.

\subsection{Galactic Centre}\label{sec:Galactic Centre}
As mentioned before, our SNR sample cannot be assumed to be unbiased near the Galactic Centre ($|l| \lesssim 10^{\circ}$). The analysis of the centre-excluded sample has ensured that any bias present near the centre does not affect our conclusions for unaffected regions, but has not reduced the bias present in that region itself. This means that any inferences about the SNR distribution at the Galactic Centre should be made very carefully. We argue, based on the discussion presented in section~\ref{sec:SNR detection biases}, that a truly complete sample of all SNRs with a surface brightness above $1.3 \times 10^{-20}$ W m$^{-2}$ Hz$^{-1}$ sr$^{-1}$ at 1 GHz would feature more remnants near the Galactic Centre than presented here. Therefore, it is noteworthy that the 'true' SNR distribution is likely more centrally concentrated than the distribution we find. This is despite the fact that we already find a more centrally concentrated distribution compared to earlier studies (see Figs.~\ref{fig:comparison_dens} and \ref{fig:comparison_num}).

With regards to the Galactic bar there are two important things to note. Firstly, the bias at the Galactic Centre effectively limits our ability to treat the bar differently from the disk, since any inferences would be highly biased. Secondly, since the bar is elliptical in the plane of the Galaxy, we would not expect a sharp cutoff between the disk and bar regions based on a purely radial SNR distribution, even if these regions would not follow the same SNR distribution.

\subsection{Interpretation/Implications}
Given their lifetime, the distribution of SNRs in the Galaxy gives an incomplete record of all SNe that occurred over the last 20 to 80 kyr. Moreover, given that SNRs are much shorter lived than massive stars and the relative rarity of type Ia compared to core-collapse SNe, the SNR distribution is expected to follow not only the core-collapse SN distribution, but also the distribution of massive stars. Multiple studies have been performed over the years investigating the distribution of massive stars and their formation regions \citep[e.g.][]{Comeron_1996, Bronfman_2000, Urquhart_2014}. Commonly, a peak is found around $\sim4-5$ kpc, with an exponential decrease towards higher Galactocentric radii. The study done by \citet{Urquhart_2014} also found other peaks in the distribution that they attribute to local structure in the disk (e.g.~spiral arms). Although \citet{Urquhart_2014} where not sensitive to Galactocentric radii $\lesssim3$ kpc, their distribution does seem to plateau if not decrease for $3\lesssim R_\textrm{gal}\lesssim5$ kpc. This seems to correspond to a local minimum in the H$_2$ surface density at $\sim2$ kpc \citep{Nakanishi_2006}. Although we see no clear evidence for these structures in our data, this could be the consequence of our use of the 1D projection in $l$.

Having discussed the Galactic Centre and massive star formation, it is also interesting to discuss the shapes of the distributions shown in Fig.~\ref{fig:comparison_dens} and Fig.~\ref{fig:comparison_num}. Given the interplay between surface area and surface density, a peak in the number of SNRs at $R_{gal}\neq 0$ is not unexpected. The peak in the surface density of SNRs found by \citet{Green_2015} and \citet{Case_1998} on the other hand would imply that the SNR formation mechanism would be most efficient at that particular radius. Given that the SNR sample around the Galactic Centre is likely incomplete, we expect that if such a peak were to be present it would be located at $R_{gal}\leq1.7$ kpc, which is the peak radius we find using a MGD. This would place it withing the Galactic bar region \citep[e.g.][]{Babusiaux_2005}. We expect the presence of this peak to be unlikely, since there is a lack of observational evidence supporting it and no clear interpretation for a peak in this region. It could therefore be an artifact of the use of a MGD to describe the radial SNR distribution.

As we noted earlier, SNe are the dominant source of chemical enrichment of galaxies with heavy elements. For this reason a comparison with the radial metallicity distribution in the Galaxy is quite natural. However, given the complex formational history of the Galaxy, one of course needs to be careful in direct comparisons. By studying Cepheids with $3 \lesssim R_{\textrm{gal}}\lesssim15$ kpc, \citet{Luck_2011} find an exponential relation between metallicity and Galactocentric radius. Assuming that this enrichment happens on a timescale smaller or similar to that of the radial diffusion of matter in the Galaxy in combination with the assumption that the current SNR distribution is representative for that in the past this result seems to support an exponential SNR distribution for $R_{\textrm{gal}}\gtrsim3-5$ kpc.

\subsection{Future work}
It would be interesting to see how well the SNR distribution in other spiral galaxies is described by the exponential profile in comparison with the other models. In this way it could be investigated if, and in how far, the distribution found here holds a universal power in describing SNR distributions. What would also make this advantageous is that the full spacial $\Sigma-D$ relation that causes a large distance determination scatter in the Galaxy is not needed for extra-galactic sources. Especially in face-on spiral galaxies, all remnants are at approximately the same distance in a 2D configuration. The distance to the centre of their host galaxy can then be determined using geometry. Unfortunately it would be difficult to compare the results with what has been done here. Extra-galactic SNRs suffer from different detection biases that would change per host galaxy. Quantifying these biases would therefore be another study in itself. Since comparing the distributions directly as if no biases are present would be of questionable value, we have opted to focus solely on the Galactic distribution instead. 

With regards to discovering new SNRs, the region surrounding the Galactic Centre ($|l|\lesssim10^{\circ}$) is the most promising. Although the observational challenges are greatest in this region, it is also expected to harbor the greatest undiscovered Galactic SNR density per square unit area on the sky.

\section{Conclusions}\label{sec:conclusion}
By imposing a surface-brightness-completeness limit of $1.3 \times 10^{-20}$ W m$^{-2}$ Hz$^{-1}$ sr$^{-1}$ at 1 GHz we have analysed the radial distribution of SNRs in the Galaxy. We fitted multiple model distributions to determine if the use of a MGD model, that has seen persistent use in the literature, is really justified. The analysis has been done for two samples. The first sample included every SNR above the surface-brightness threshold and the second sample only included, from these remaining SNRs, the ones with $l>\pm10^{\circ}$, since biases are mainly present near the Galactic Centre. In neither of these samples does the MGD provide the best fitting results. Moreover, the MGD does not provide significantly better fits than a simple exponential profile in either of these samples. We find that a simple exponential profile of the form given in equation~(\ref{eq:exp}), with $\beta=2.46^{+0.39}_{-0.33}$ and an integrated SNR density peak at 3.3 kpc, is the most consistent and, with one free parameter, least complicated model. 

Given the relatively high bias in the data set around the Galactic Centre, no strict inferences have been made about the central region of the Galaxy. By comparing our findings to those of studies investigating the radial distribution of massive star formation and metallicity, we find indirect evidence supporting an exponential SNR surface density for $R_{\textrm{gal}}\gtrsim3-5$ kpc. Our study, therefore, suggests that preference should be given to a simple exponential distribution over the MGD given in equation~(\ref{eq:gamma_distribution}) when describing the radial SNR distribution in the Galaxy.

\section*{Acknowledgements}
We would like to give special thanks to Ciaran Rogers and Stefanie Fijma for their valuable feedback on an earlier version of this paper.

\section*{Data heading}
The data underlying this article are available at \url{https://doi.org/10.1007/s12036-019-9601-6}.





\bibliographystyle{mnras}
\bibliography{references} 

\begin{thebibliography}{}
\makeatletter
\relax
\def\mn@urlcharsother{\let\do\@makeother \do\$\do\&\do\#\do\^\do\_\do\%\do\~}
\def\mn@doi{\begingroup\mn@urlcharsother \@ifnextchar [ {\mn@doi@}
  {\mn@doi@[]}}
\def\mn@doi@[#1]#2{\def\@tempa{#1}\ifx\@tempa\@empty \href
  {http://dx.doi.org/#2} {doi:#2}\else \href {http://dx.doi.org/#2} {#1}\fi
  \endgroup}
\def\mn@eprint#1#2{\mn@eprint@#1:#2::\@nil}
\def\mn@eprint@arXiv#1{\href {http://arxiv.org/abs/#1} {{\tt arXiv:#1}}}
\def\mn@eprint@dblp#1{\href {http://dblp.uni-trier.de/rec/bibtex/#1.xml}
  {dblp:#1}}
\def\mn@eprint@#1:#2:#3:#4\@nil{\def\@tempa {#1}\def\@tempb {#2}\def\@tempc
  {#3}\ifx \@tempc \@empty \let \@tempc \@tempb \let \@tempb \@tempa \fi \ifx
  \@tempb \@empty \def\@tempb {arXiv}\fi \@ifundefined
  {mn@eprint@\@tempb}{\@tempb:\@tempc}{\expandafter \expandafter \csname
  mn@eprint@\@tempb\endcsname \expandafter{\@tempc}}}

\bibitem[\protect\citeauthoryear{Babusiaux \& Gilmore}{Babusiaux \&
  Gilmore}{2005}]{Babusiaux_2005}
Babusiaux C.,  Gilmore G.,  2005, \mn@doi [Monthly Notices of the Royal
  Astronomical Society] {10.1111/j.1365-2966.2005.08828.x}, 358, 1309

\bibitem[\protect\citeauthoryear{{Bell}, {Schure}, {Reville}  \&
  {Giacinti}}{{Bell} et~al.}{2013}]{Bell_2013}
{Bell} A.~R.,  {Schure} K.~M.,  {Reville} B.,   {Giacinti} G.,  2013, \mn@doi
  [\mnras] {10.1093/mnras/stt179}, \href
  {https://ui.adsabs.harvard.edu/abs/2013MNRAS.431..415B} {431, 415}

\bibitem[\protect\citeauthoryear{{Berezhko}, {P{\"u}hlhofer}  \&
  {V{\"o}lk}}{{Berezhko} et~al.}{2003}]{Berezhko_2003}
{Berezhko} E.~G.,  {P{\"u}hlhofer} G.,   {V{\"o}lk} H.~J.,  2003, \mn@doi
  [\aap] {10.1051/0004-6361:20030033}, \href
  {https://ui.adsabs.harvard.edu/abs/2003A&A...400..971B} {400, 971}

\bibitem[\protect\citeauthoryear{{Bethe}}{{Bethe}}{1990}]{Bethe_1990}
{Bethe} H.~A.,  1990, \mn@doi [Reviews of Modern Physics]
  {10.1103/RevModPhys.62.801}, \href
  {https://ui.adsabs.harvard.edu/abs/1990RvMP...62..801B} {62, 801}

\bibitem[\protect\citeauthoryear{{Blasi}}{{Blasi}}{2011}]{Blasi_2011}
{Blasi} P.,  2011, in {Giani} S.,  {Leroy} C.,   {Rancoita} P.~G.,  eds, Cosmic
  Rays for Particle and Astroparticle Physics. pp 493--506 (\mn@eprint {arXiv}
  {1012.5005}), \mn@doi{10.1142/9789814329033_0061}

\bibitem[\protect\citeauthoryear{{Blondin}, {Wright}, {Borkowski}  \&
  {Reynolds}}{{Blondin} et~al.}{1998}]{Blondin_1998}
{Blondin} J.~M.,  {Wright} E.~B.,  {Borkowski} K.~J.,   {Reynolds} S.~P.,
  1998, \mn@doi [\apj] {10.1086/305708}, \href
  {https://ui.adsabs.harvard.edu/abs/1998ApJ...500..342B} {500, 342}

\bibitem[\protect\citeauthoryear{{Bronfman}, {Casassus}, {May}  \&
  {Nyman}}{{Bronfman} et~al.}{2000}]{Bronfman_2000}
{Bronfman} L.,  {Casassus} S.,  {May} J.,   {Nyman} L.~{\r{A}}.,  2000, \aap,
  \href {https://ui.adsabs.harvard.edu/abs/2000A&A...358..521B} {358, 521}

\bibitem[\protect\citeauthoryear{{Case} \& {Bhattacharya}}{{Case} \&
  {Bhattacharya}}{1998}]{Case_1998}
{Case} G.~L.,  {Bhattacharya} D.,  1998, \mn@doi [\apj] {10.1086/306089}, \href
  {https://ui.adsabs.harvard.edu/abs/1998ApJ...504..761C} {504, 761}

\bibitem[\protect\citeauthoryear{{Comeron} \& {Torra}}{{Comeron} \&
  {Torra}}{1996}]{Comeron_1996}
{Comeron} F.,  {Torra} J.,  1996, \aap, \href
  {https://ui.adsabs.harvard.edu/abs/1996A&A...314..776C} {314, 776}

\bibitem[\protect\citeauthoryear{{Driver} et~al.,}{{Driver}
  et~al.}{2006}]{Driver_2006}
{Driver} S.~P.,  et~al., 2006, \mn@doi [\mnras]
  {10.1111/j.1365-2966.2006.10126.x}, \href
  {https://ui.adsabs.harvard.edu/abs/2006MNRAS.368..414D} {368, 414}

\bibitem[\protect\citeauthoryear{Genzel, Eisenhauer  \& Gillessen}{Genzel
  et~al.}{2010}]{Genzel_2010}
Genzel R.,  Eisenhauer F.,   Gillessen S.,  2010, \mn@doi [Reviews of Modern
  Physics] {10.1103/revmodphys.82.3121}, 82, 3121

\bibitem[\protect\citeauthoryear{{Graham} \& {Driver}}{{Graham} \&
  {Driver}}{2005}]{Graham_2005}
{Graham} A.~W.,  {Driver} S.~P.,  2005, \mn@doi [\pasa] {10.1071/AS05001},
  \href {https://ui.adsabs.harvard.edu/abs/2005PASA...22..118G} {22, 118}

\bibitem[\protect\citeauthoryear{{Graur}, {Bianco}, {Modjaz}, {Shivvers},
  {Filippenko}, {Li}  \& {Smith}}{{Graur} et~al.}{2017}]{Graur_2017}
{Graur} O.,  {Bianco} F.~B.,  {Modjaz} M.,  {Shivvers} I.,  {Filippenko} A.~V.,
   {Li} W.,   {Smith} N.,  2017, \mn@doi [\apj] {10.3847/1538-4357/aa5eb7},
  \href {https://ui.adsabs.harvard.edu/abs/2017ApJ...837..121G} {837, 121}

\bibitem[\protect\citeauthoryear{{Green}}{{Green}}{2004}]{Green_2004}
{Green} D.~A.,  2004, Bulletin of the Astronomical Society of India, \href
  {https://ui.adsabs.harvard.edu/abs/2004BASI...32..335G} {32, 335}

\bibitem[\protect\citeauthoryear{{Green}}{{Green}}{2015}]{Green_2015}
{Green} D.~A.,  2015, \mn@doi [\mnras] {10.1093/mnras/stv1885}, \href
  {https://ui.adsabs.harvard.edu/abs/2015MNRAS.454.1517G} {454, 1517}

\bibitem[\protect\citeauthoryear{{Green}}{{Green}}{2019}]{Green_2019}
{Green} D.~A.,  2019, \mn@doi [Journal of Astrophysics and Astronomy]
  {10.1007/s12036-019-9601-6}, \href
  {https://ui.adsabs.harvard.edu/abs/2019JApA...40...36G} {40, 36}

\bibitem[\protect\citeauthoryear{{Haslam}, {Salter}, {Stoffel}  \&
  {Wilson}}{{Haslam} et~al.}{1982}]{Haslam_1982}
{Haslam} C.~G.~T.,  {Salter} C.~J.,  {Stoffel} H.,   {Wilson} W.~E.,  1982,
  \aaps, \href {https://ui.adsabs.harvard.edu/abs/1982A&AS...47....1H} {47, 1}

\bibitem[\protect\citeauthoryear{{Hillebrandt} \& {Niemeyer}}{{Hillebrandt} \&
  {Niemeyer}}{2000}]{Hillebrandt_2000}
{Hillebrandt} W.,  {Niemeyer} J.~C.,  2000, \mn@doi [\araa]
  {10.1146/annurev.astro.38.1.191}, \href
  {https://ui.adsabs.harvard.edu/abs/2000ARA&A..38..191H} {38, 191}

\bibitem[\protect\citeauthoryear{Kennea et~al.,}{Kennea
  et~al.}{2013}]{Kennea_2013}
Kennea J.~A.,  et~al., 2013, \mn@doi [The Astrophysical Journal]
  {10.1088/2041-8205/770/2/l24}, 770, L24

\bibitem[\protect\citeauthoryear{{Li}, {Chornock}, {Leaman}, {Filippenko},
  {Poznanski}, {Wang}, {Ganeshalingam}  \& {Mannucci}}{{Li}
  et~al.}{2011}]{Li_2011}
{Li} W.,  {Chornock} R.,  {Leaman} J.,  {Filippenko} A.~V.,  {Poznanski} D.,
  {Wang} X.,  {Ganeshalingam} M.,   {Mannucci} F.,  2011, \mn@doi [\mnras]
  {10.1111/j.1365-2966.2011.18162.x}, \href
  {https://ui.adsabs.harvard.edu/abs/2011MNRAS.412.1473L} {412, 1473}

\bibitem[\protect\citeauthoryear{{Luck} \& {Lambert}}{{Luck} \&
  {Lambert}}{2011}]{Luck_2011}
{Luck} R.~E.,  {Lambert} D.~L.,  2011, \mn@doi [\aj]
  {10.1088/0004-6256/142/4/136}, \href
  {https://ui.adsabs.harvard.edu/abs/2011AJ....142..136L} {142, 136}

\bibitem[\protect\citeauthoryear{{Lucy}}{{Lucy}}{2000}]{Lucy_2000}
{Lucy} L.~B.,  2000, \mn@doi [\mnras] {10.1046/j.1365-8711.2000.03655.x}, \href
  {https://ui.adsabs.harvard.edu/abs/2000MNRAS.318...92L} {318, 92}

\bibitem[\protect\citeauthoryear{{Maeda} et~al.,}{{Maeda}
  et~al.}{2002}]{Maeda_2002}
{Maeda} Y.,  et~al., 2002, \mn@doi [\apj] {10.1086/339773}, \href
  {https://ui.adsabs.harvard.edu/abs/2002ApJ...570..671M} {570, 671}

\bibitem[\protect\citeauthoryear{{Maoz}, {Mannucci}  \& {Nelemans}}{{Maoz}
  et~al.}{2014}]{Moaz_2014}
{Maoz} D.,  {Mannucci} F.,   {Nelemans} G.,  2014, \mn@doi [\araa]
  {10.1146/annurev-astro-082812-141031}, \href
  {https://ui.adsabs.harvard.edu/abs/2014ARA&A..52..107M} {52, 107}

\bibitem[\protect\citeauthoryear{Nakanishi \& Sofue}{Nakanishi \&
  Sofue}{2006}]{Nakanishi_2006}
Nakanishi H.,  Sofue Y.,  2006, \mn@doi [Publications of the Astronomical
  Society of Japan] {10.1093/pasj/58.5.847}, 58, 847

\bibitem[\protect\citeauthoryear{{Platania}, {Bensadoun}, {Bersanelli}, {De
  Amici}, {Kogut}, {Levin}, {Maino}  \& {Smoot}}{{Platania}
  et~al.}{1998}]{Platania_1998}
{Platania} P.,  {Bensadoun} M.,  {Bersanelli} M.,  {De Amici} G.,  {Kogut} A.,
  {Levin} S.,  {Maino} D.,   {Smoot} G.~F.,  1998, \mn@doi [\apj]
  {10.1086/306175}, \href
  {https://ui.adsabs.harvard.edu/abs/1998ApJ...505..473P} {505, 473}

\bibitem[\protect\citeauthoryear{{Sarbadhicary}, {Badenes}, {Chomiuk},
  {Caprioli}  \& {Huizenga}}{{Sarbadhicary} et~al.}{2017}]{Sarbadhicary_2017}
{Sarbadhicary} S.~K.,  {Badenes} C.,  {Chomiuk} L.,  {Caprioli} D.,
  {Huizenga} D.,  2017, \mn@doi [\mnras] {10.1093/mnras/stw2566}, \href
  {https://ui.adsabs.harvard.edu/abs/2017MNRAS.464.2326S} {464, 2326}

\bibitem[\protect\citeauthoryear{{S{\'e}rsic}}{{S{\'e}rsic}}{1963}]{Sersic_1963}
{S{\'e}rsic} J.~L.,  1963, Boletin de la Asociacion Argentina de Astronomia La
  Plata Argentina, \href
  {https://ui.adsabs.harvard.edu/abs/1963BAAA....6...41S} {6, 41}

\bibitem[\protect\citeauthoryear{{Stecker} \& {Jones}}{{Stecker} \&
  {Jones}}{1977}]{Stecker_1977}
{Stecker} F.~W.,  {Jones} F.~C.,  1977, Technical report, {The galactic halo
  question: New size constraints from galactic gamma-ray data}

\bibitem[\protect\citeauthoryear{{Tammann}, {Loeffler}  \&
  {Schroeder}}{{Tammann} et~al.}{1994}]{Tammann_1994}
{Tammann} G.~A.,  {Loeffler} W.,   {Schroeder} A.,  1994, \mn@doi [\apjs]
  {10.1086/192002}, \href
  {https://ui.adsabs.harvard.edu/abs/1994ApJS...92..487T} {92, 487}

\bibitem[\protect\citeauthoryear{{Urquhart}, {Figura}, {Moore}, {Hoare},
  {Lumsden}, {Mottram}, {Thompson}  \& {Oudmaijer}}{{Urquhart}
  et~al.}{2014}]{Urquhart_2014}
{Urquhart} J.~S.,  {Figura} C.~C.,  {Moore} T.~J.~T.,  {Hoare} M.~G.,
  {Lumsden} S.~L.,  {Mottram} J.~C.,  {Thompson} M.~A.,   {Oudmaijer} R.~D.,
  2014, \mn@doi [\mnras] {10.1093/mnras/stt2006}, \href
  {https://ui.adsabs.harvard.edu/abs/2014MNRAS.437.1791U} {437, 1791}

\bibitem[\protect\citeauthoryear{{Wheaton}, {Dunklee}, {Jacobsen}, {Ling},
  {Mahoney}  \& {Radocinski}}{{Wheaton} et~al.}{1995}]{Wheaton_1995}
{Wheaton} W.~A.,  {Dunklee} A.~L.,  {Jacobsen} A.~S.,  {Ling} J.~C.,  {Mahoney}
  W.~A.,   {Radocinski} R.~G.,  1995, \mn@doi [\apj] {10.1086/175077}, \href
  {https://ui.adsabs.harvard.edu/abs/1995ApJ...438..322W} {438, 322}

\bibitem[\protect\citeauthoryear{Widrow, Pym  \& Dubinski}{Widrow
  et~al.}{2008}]{Widrow_2008}
Widrow L.~M.,  Pym B.,   Dubinski J.,  2008, \mn@doi [The Astrophysical
  Journal] {10.1086/587636}, 679, 1239

\bibitem[\protect\citeauthoryear{{Woosley} \& {Weaver}}{{Woosley} \&
  {Weaver}}{1995}]{Woosley_1995}
{Woosley} S.~E.,  {Weaver} T.~A.,  1995, \mn@doi [\apjs] {10.1086/192237},
  \href {https://ui.adsabs.harvard.edu/abs/1995ApJS..101..181W} {101, 181}

\bibitem[\protect\citeauthoryear{{de Vaucouleurs}}{{de
  Vaucouleurs}}{1948}]{deVaucouleurs_1948}
{de Vaucouleurs} G.,  1948, Annales d'Astrophysique, \href
  {https://ui.adsabs.harvard.edu/abs/1948AnAp...11..247D} {11, 247}

\makeatother
\end{thebibliography}




\appendix

\section{Model fits}\label{sec:Plots}
Here we provide all the model fits that have been performed with the different distributions. Both the model fits to the complete and centre-excluded data sets are shown.

\begin{figure*}
\centering
\begin{tabular}{cc}
\includegraphics[width = 0.47\textwidth]{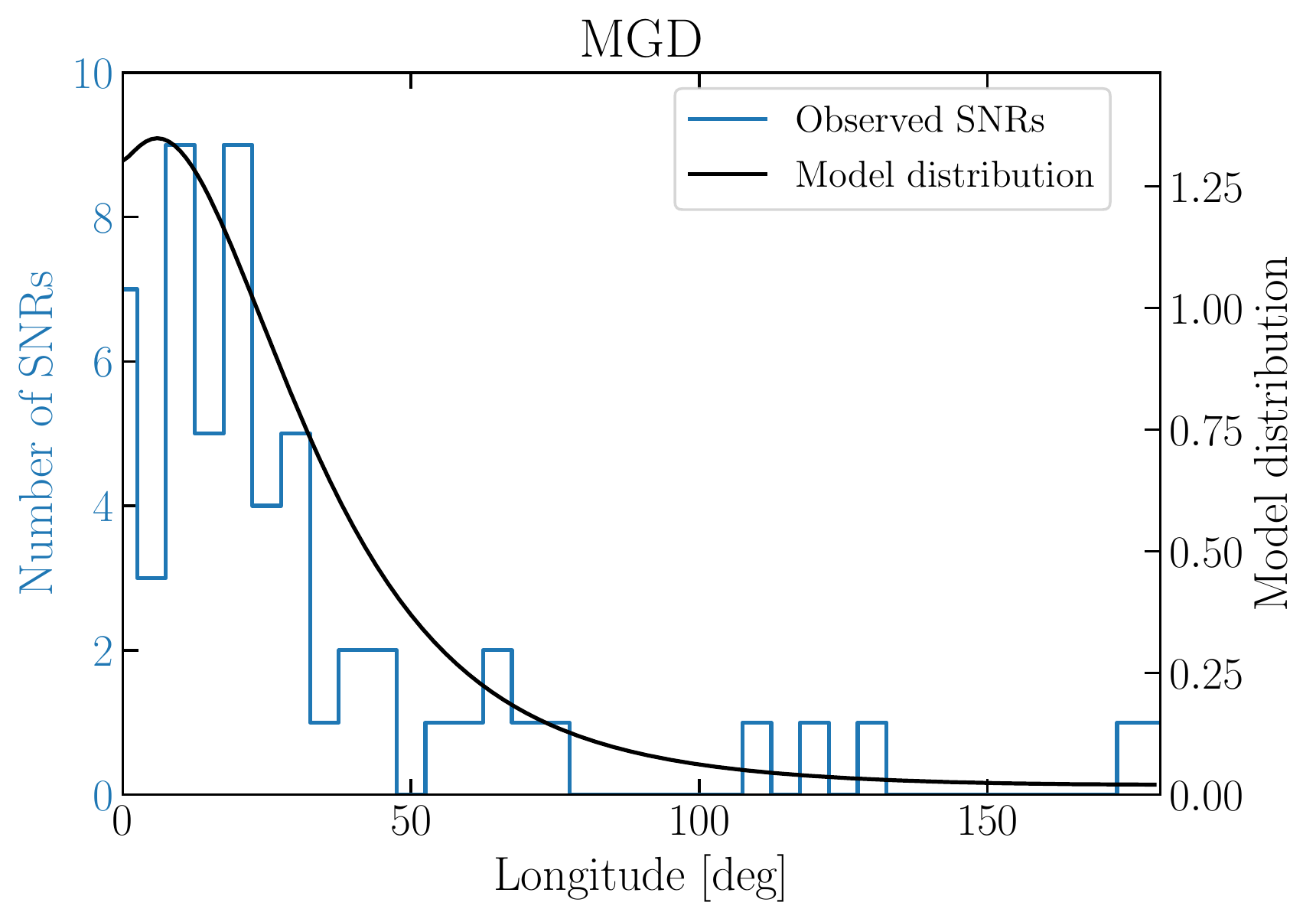} &
\includegraphics[width = 0.47\textwidth]{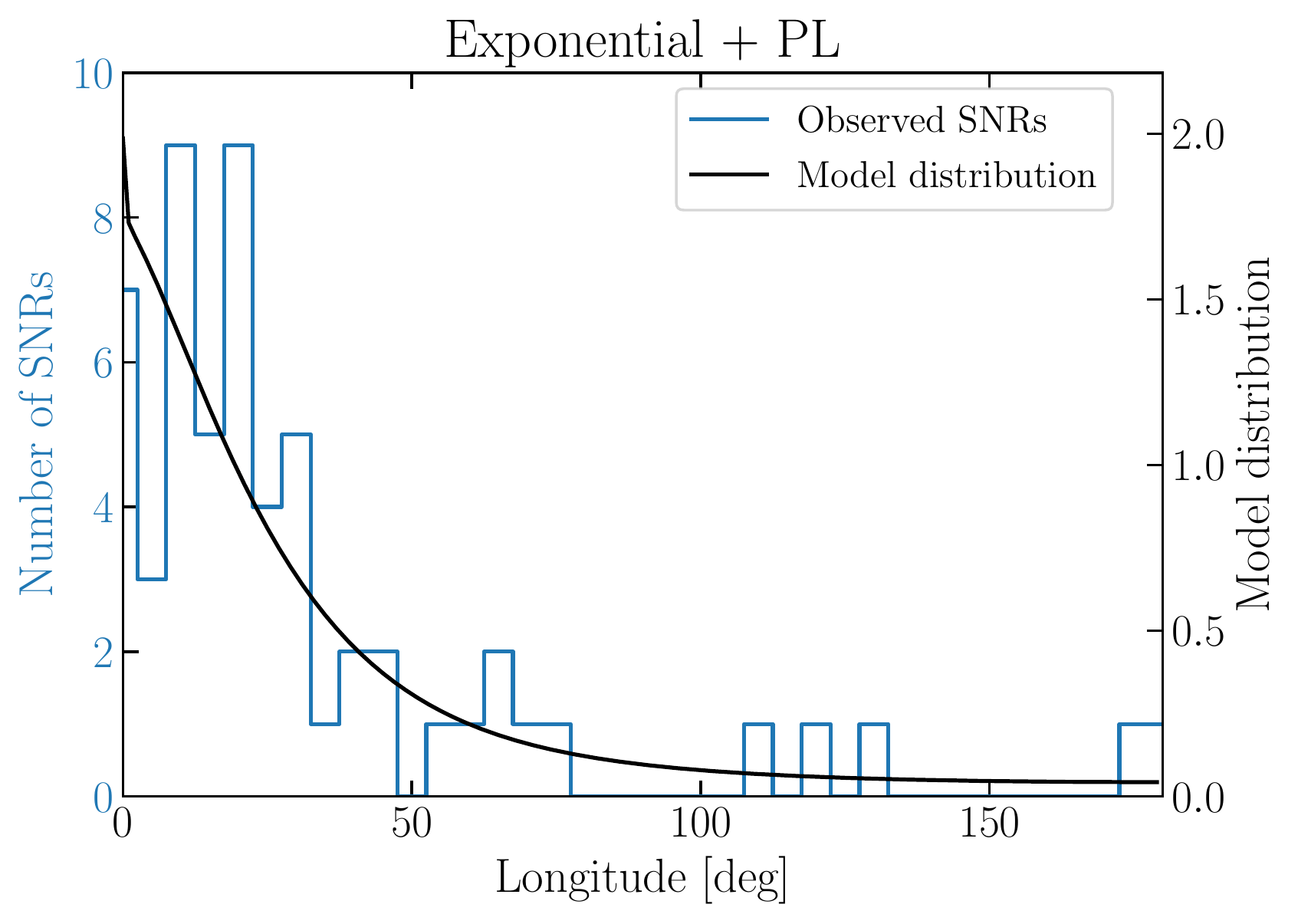} \\
\includegraphics[width = 0.47\textwidth]{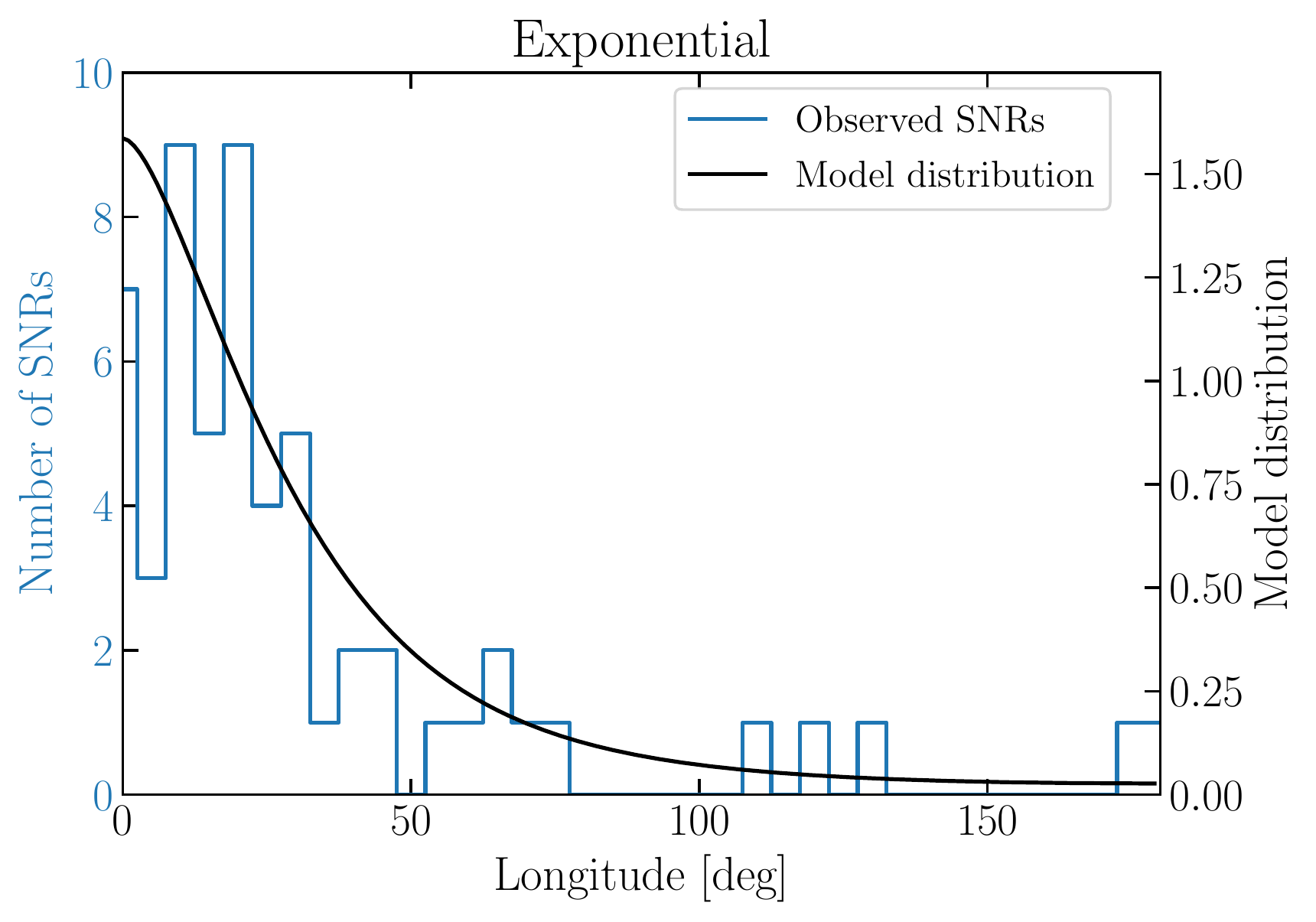} &
\includegraphics[width = 0.47\textwidth]{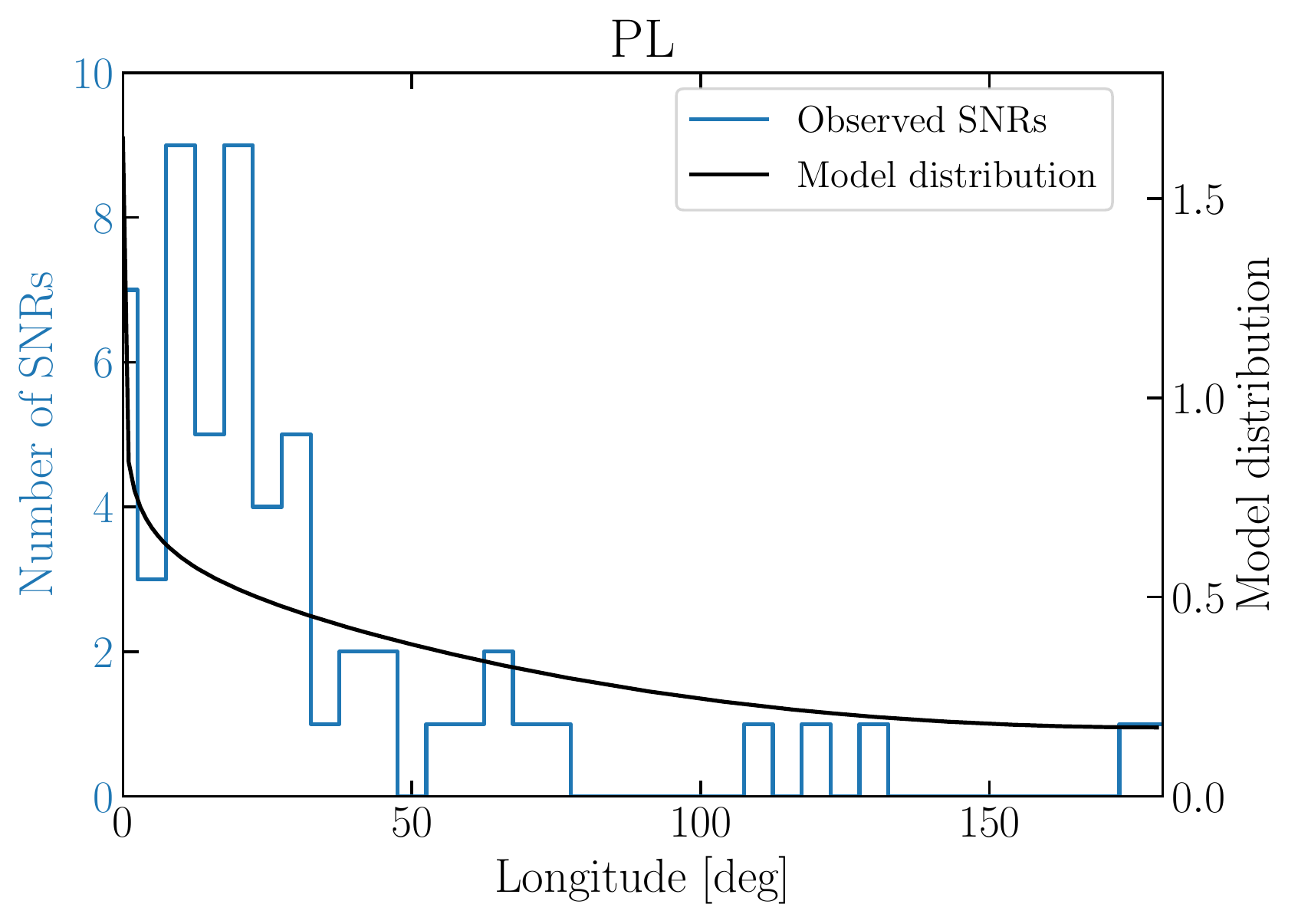} \\
\includegraphics[width = 0.47\textwidth]{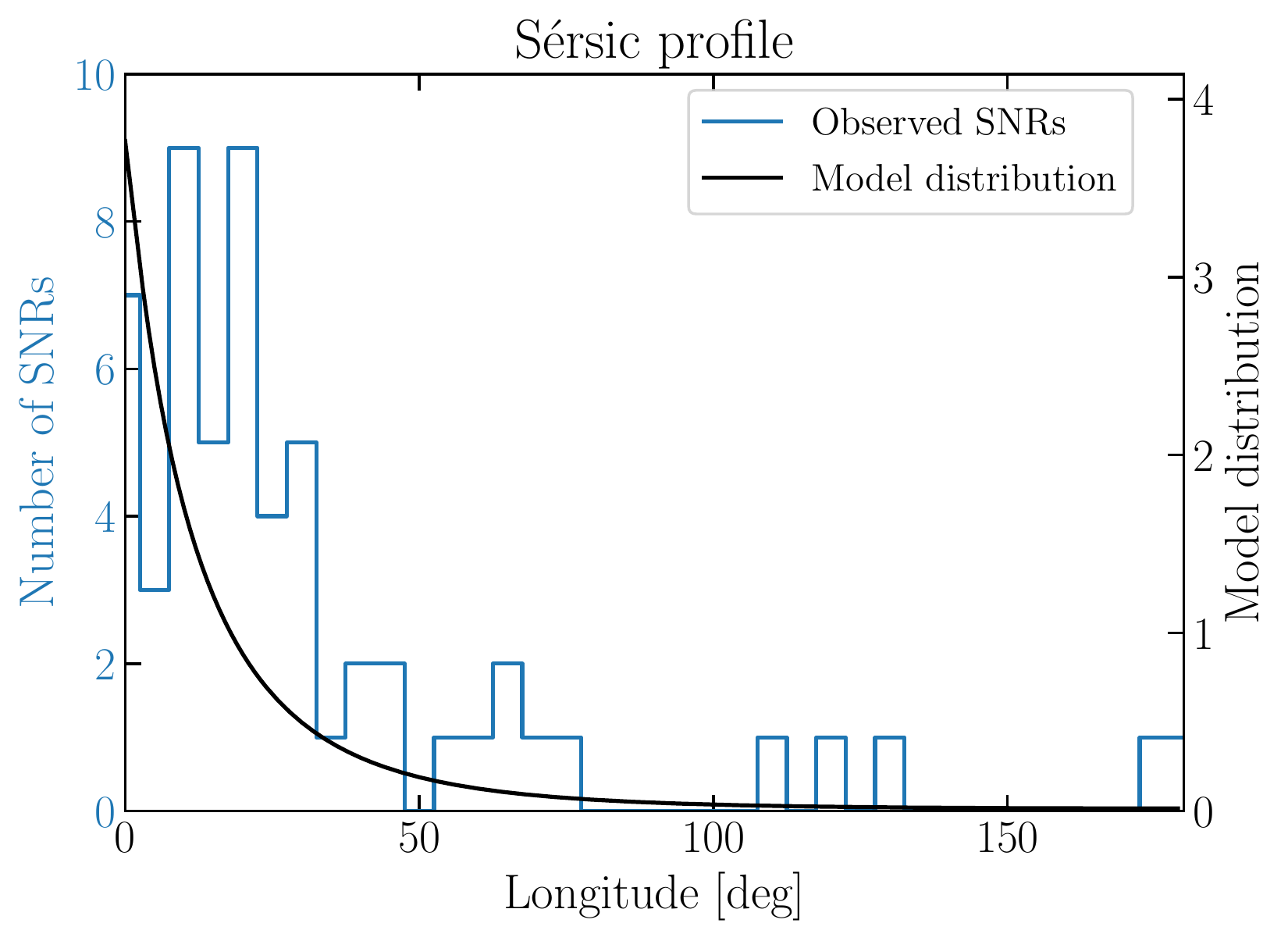} \\
\end{tabular}
\caption{The best model fits to the complete surface-brightness-limited data set.}
\label{fig:compl}
\end{figure*}
\newpage
\begin{figure*}
\centering
\begin{tabular}{cc}
\includegraphics[width = 0.47\textwidth]{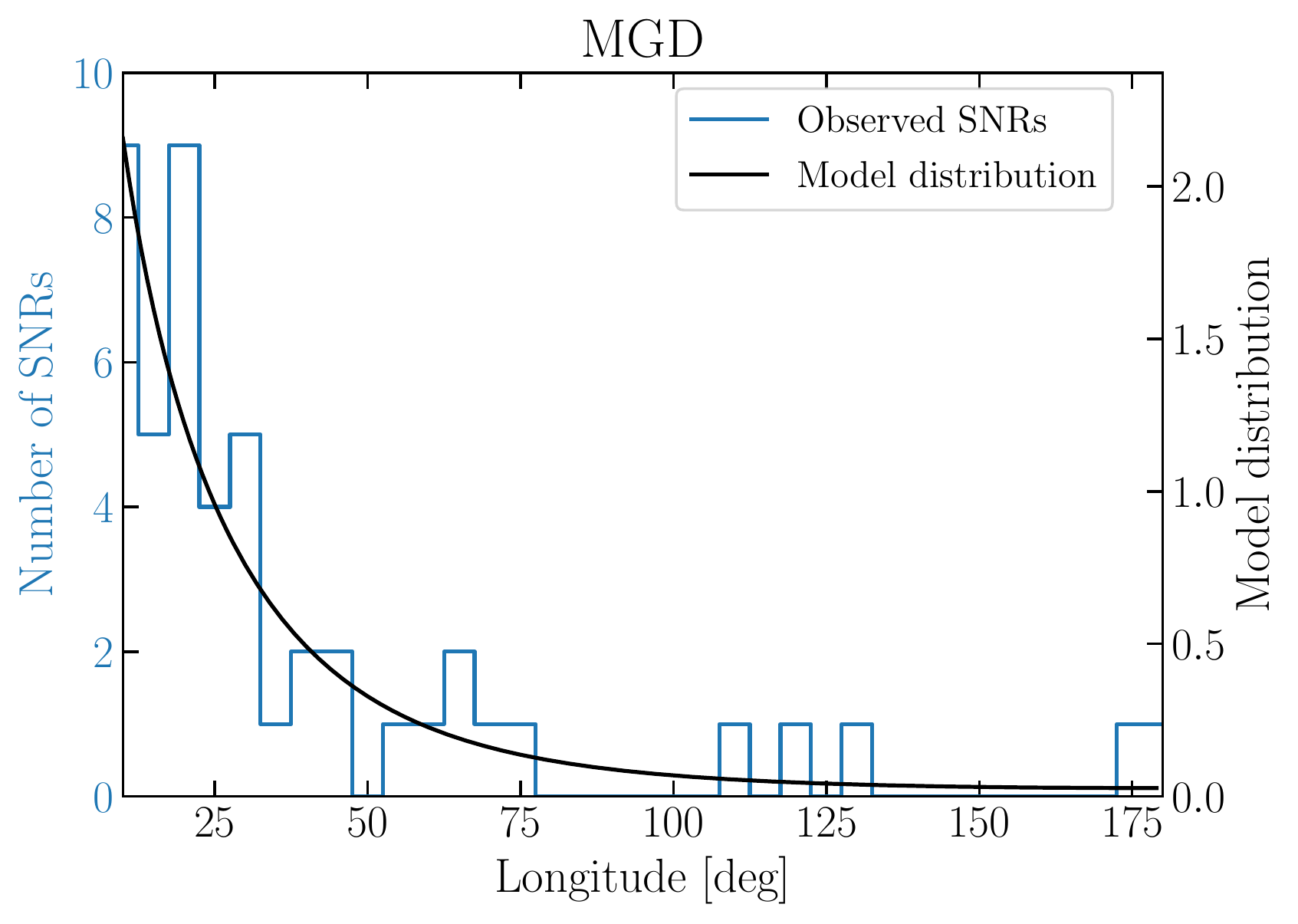} &
\includegraphics[width = 0.47\textwidth]{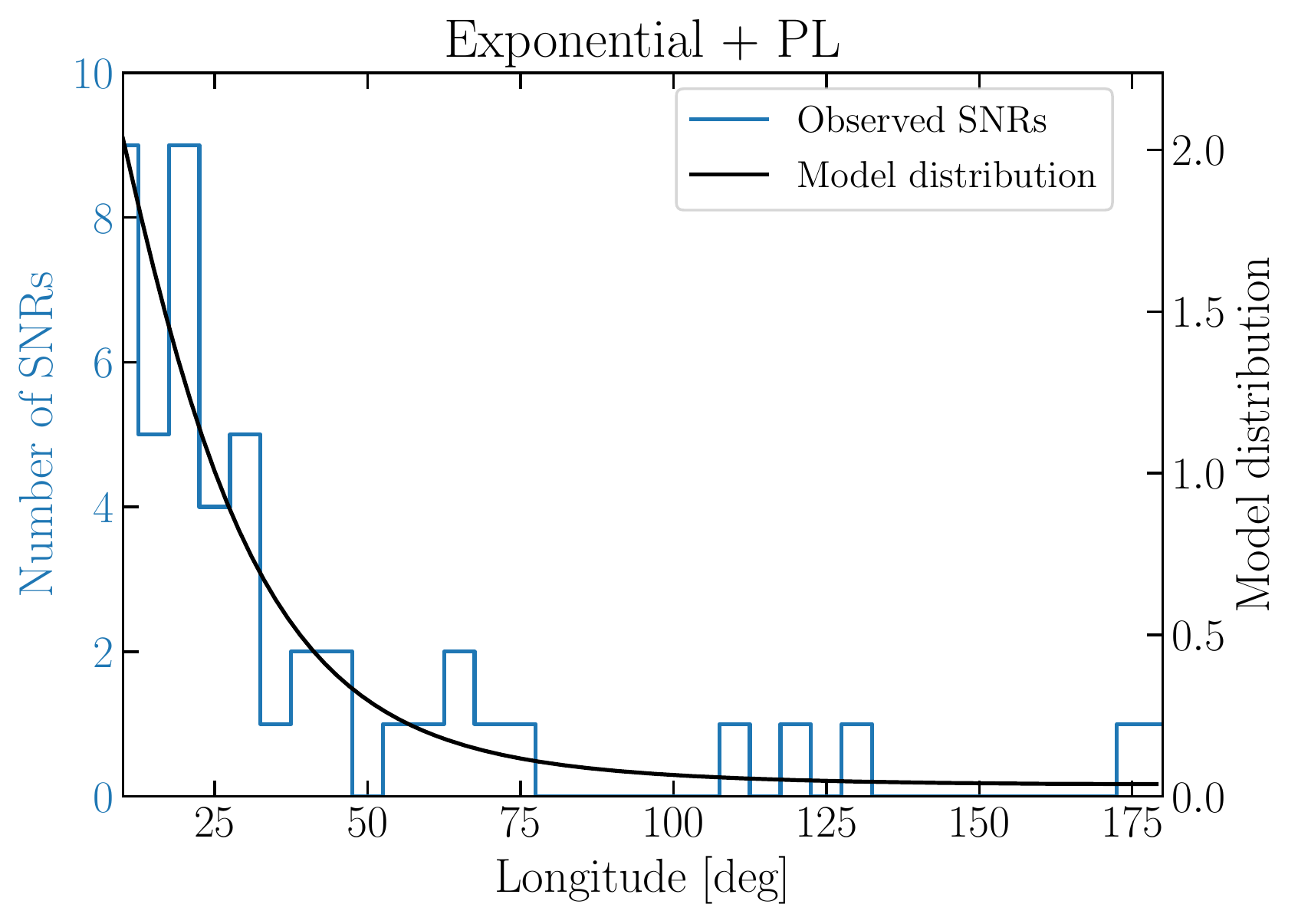} \\
\includegraphics[width = 0.47\textwidth]{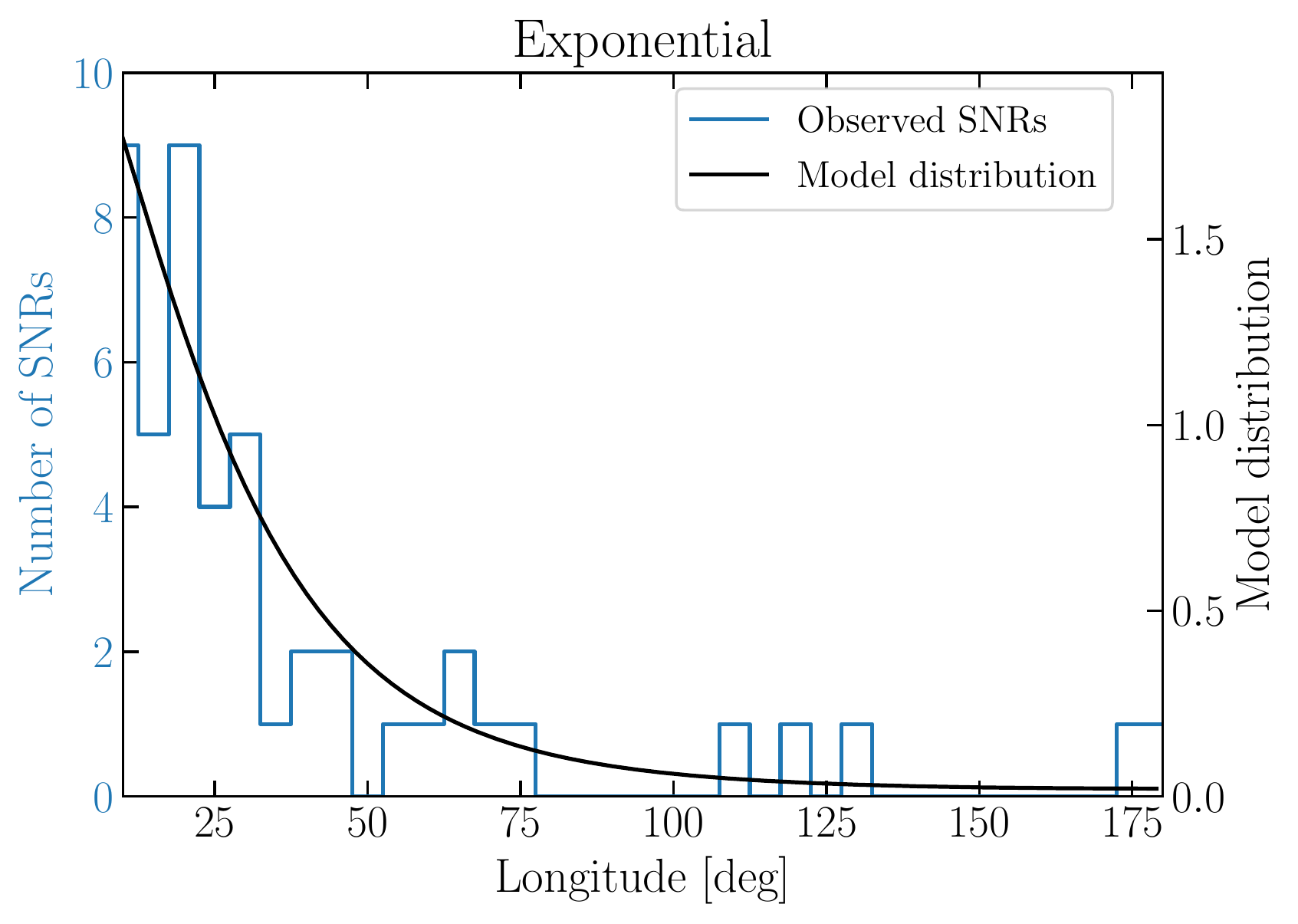} &
\includegraphics[width = 0.47\textwidth]{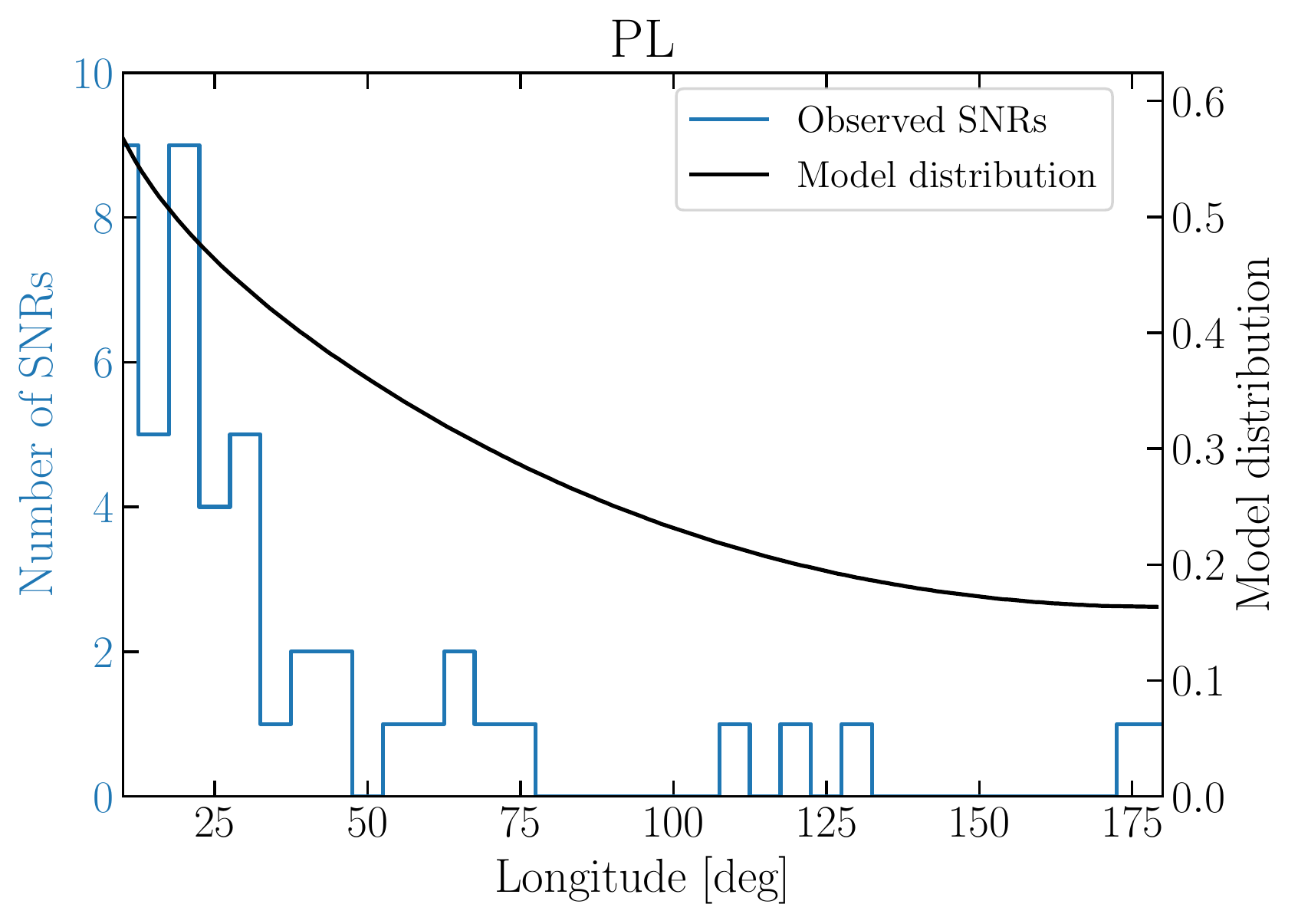} \\
\includegraphics[width = 0.47\textwidth]{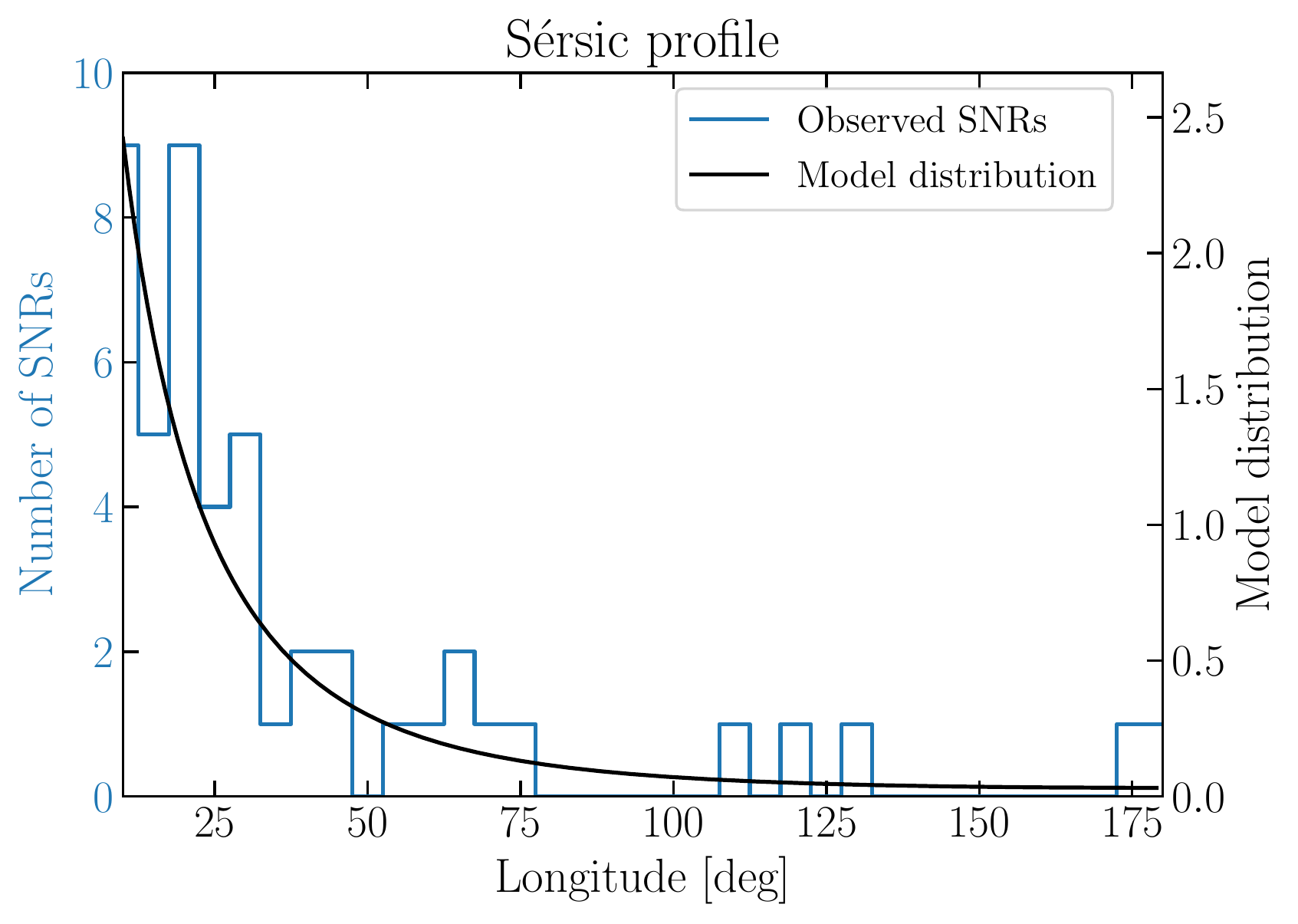} \\
\end{tabular}
\caption{The best model fits to the centre-excluded sample.}
\end{figure*}


\bsp	
\label{lastpage}
\end{document}